\documentclass[sigconf, nonacm]{acmart}

\usepackage{amsmath}
\usepackage{mathtools}
\usepackage{enumitem}
\usepackage{textcomp}
\usepackage{graphicx}
\usepackage{multirow}
\usepackage{adjustbox}
\usepackage{booktabs}
\usepackage{xcolor}
\usepackage{makecell}
\usepackage{array}
\usepackage{tcolorbox} 
\usepackage[normalem]{ulem} 
\usepackage{xcolor} 
\usepackage{tikz}
\newcommand*\filledcircled[1]{\tikz[baseline=(char.base)]{
  \node[shape=circle,fill=black,text=white,inner sep=1pt] (char) {\sffamily\bfseries\small #1};}}
\AtBeginDocument{%
  }

\newcommand\vldbdoi{XX.XX/XXX.XX}
\newcommand\vldbpages{XXX-XXX}
\newcommand\vldbvolume{20}
\newcommand\vldbissue{3}
\newcommand\vldbyear{2026}
\newcommand\vldbauthors{\authors}
\newcommand\vldbtitle{\shorttitle} 
\newcommand\vldbavailabilityurl{https://github.com/yoon123seul/HVMap.git}
\newcommand\vldbpagestyle{empty}

\newcommand{\ra}[1]{\textcolor{black!80!black}{#1}}
\newcommand{\rb}[1]{\textcolor{black!80!black}{#1}}

\usepackage{xcolor}
\usepackage{todonotes}

\begin{document}
\title{Federated Vector Retrieval across Heterogeneous VectorDBs}

\settopmatter{authorsperrow=4}

\author{Mintaek Lim}
\affiliation{%
  \institution{Seoul National University}
  \city{Seoul}
  \state{South Korea}
}
\email{victorlim@snu.ac.kr}

\author{Sieun Hyeon}
\affiliation{
  \institution{Seoul National University}
  \city{Seoul}
  \state{South Korea}
}
\email{zxc2692@snu.ac.kr}

\author{Geon Choi}
\affiliation{
  \institution{Seoul National University}
  \city{Seoul}
  \state{South Korea}
}
\email{cds06126@snu.ac.kr}

\author{Jaeyoung Do}
\authornote{Corresponding author.}
\affiliation{%
  \institution{Seoul National University}
  \city{Seoul}
  \country{South Korea}
}
\email{jaeyoung.do@snu.ac.kr}

\begin{abstract}

Vector databases are now core infrastructure for AI retrieval systems, yet many real deployments must serve queries over multiple independently constructed vectorDBs built with different embedding models. This creates a systems problem of federated retrieval over heterogeneous vectorDBs: vectors and similarity scores are not directly comparable across embedding spaces, while fully resolving this incompatibility through corpus-wide re-embedding and re-indexing is often operationally prohibitive. Existing methods therefore face a difficult trade-off between retrieval quality, construction cost, and query throughput. We present HVMap, a lightweight compatibility layer for federated retrieval across heterogeneous vectorDBs. Instead of rebuilding databases into a unified space, HVMap learns bidirectional mappings between embedding spaces using a fuzzy contrastive loss, enabling calibrated cross-database scoring without corpus migration or index reconstruction. HVMap is system-agnostic and further scales to multi-vectorDB settings while preserving ranking fidelity through a hub-based architecture. We evaluate HVMap in open-domain, cross-domain, and scale-out settings, and also construct a new benchmark for heterogeneous cross-domain federation. Across these settings, HVMap consistently achieves a better quality--cost--throughput trade-off than baselines, achieving up to \ra{3.59$\times$} higher nDCG@10, up to 2.01$\times$ higher query throughput, and construction time as low as 0.3\% of baseline methods. These results show that HVMap makes heterogeneous vectorDB federation practical at scale.
\end{abstract}

\maketitle

\pagestyle{\vldbpagestyle}
\begingroup\small\noindent\raggedright\textbf{PVLDB Reference Format:}\\
\vldbauthors. \vldbtitle. PVLDB, \vldbvolume(\vldbissue): \vldbpages, \vldbyear.\\
\href{https://doi.org/\vldbdoi}{doi:\vldbdoi}
\endgroup
\begingroup
\renewcommand\thefootnote{}\footnote{\noindent
This work is licensed under the Creative Commons BY-NC-ND 4.0 International License. Visit \url{https://creativecommons.org/licenses/by-nc-nd/4.0/} to view a copy of this license. For any use beyond those covered by this license, obtain permission by emailing \href{mailto:info@vldb.org}{info@vldb.org}. Copyright is held by the owner/author(s). Publication rights licensed to the VLDB Endowment. \\
\raggedright Proceedings of the VLDB Endowment, Vol. \vldbvolume, No. \vldbissue\ %
ISSN 2150-8097. \\
\href{https://doi.org/\vldbdoi}{doi:\vldbdoi} \\
}\addtocounter{footnote}{-1}\endgroup

\ifdefempty{\vldbavailabilityurl}{}{
\vspace{.1cm}
\begingroup\small\noindent\raggedright\textbf{PVLDB Artifact Availability:}\\
The source code, data, and/or other artifacts have been made available at \url{https://github.com/yoon123seul/HVMap.git}.
\endgroup
}

\begin{figure}[t!]
  \centering
  \includegraphics[width=0.9\linewidth]{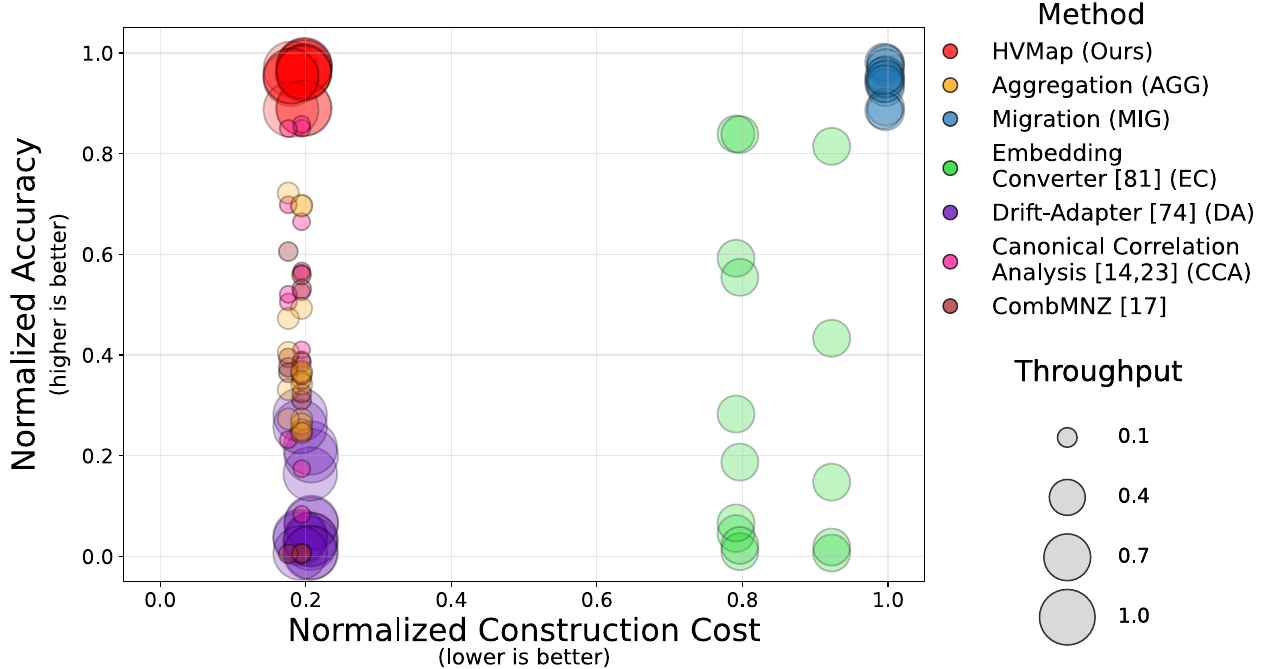}
  \caption{Accuracy--cost--throughput trade-off among methods for federated retrieval over heterogeneous vectorDBs at the 1B scale, and each bubble represents one method under a dataset--embedding-model configuration (Section~\ref{sec:expr_basecase}). Accuracy, cost, and throughput correspond to nDCG@10, construction time (hours), and query throughput (QPS), respectively. For each configuration, metrics are normalized across methods using z-score normalization followed by a sigmoid transformation, and bubble size indicates normalized query throughput.}
  \label{fig:comparison}
\end{figure}

\section{Introduction}
\label{sec:intro}

Vector databases (vectorDBs) have become a core component of modern AI data infrastructure, supporting
efficient similarity search over dense embeddings at scale. 
They are now widely used in applications such as semantic search, recommendation systems, and Retrieval-Augmented Generation (RAG) pipelines~\cite{lewis2020retrieval, balaguer2024rag, borgeaud2022improvinglanguagemodelsretrieving, izacard2022distillingknowledgereaderretriever, khandelwal2020generalizationmemorizationnearestneighbor, shi2023replugretrievalaugmentedblackboxlanguage}. 
As AI applications increasingly rely on embedding-based access to large collections of unstructured data, vectorDBs are no longer standalone components; they are long-lived system infrastructure that must remain operational even as embedding models, domains, and corpora evolve.

In practice, this evolution rarely results in a single, unified vector space.
Instead, many real-world systems maintain multiple vectorDBs built with different embedding models. 
A production deployment may preserve legacy indexes while gradually adopting a new embedding model, or simultaneously serve multiple domains, each using a specialized encoder. 
Supporting a single query over such databases naturally leads to the problem of \textit{federated retrieval across heterogeneous vectorDBs}: executing retrieval over multiple independently managed vectorDBs whose embeddings are not directly comparable. 
In many production deployments, heterogeneous vectorDBs are not an exception but an operational consequence of model upgrades, domain specialization, and incremental migration. 
A system design that assumes immediate corpus-wide re-embedding is therefore often impractical.

This setting arises naturally in at least two common scenarios.
First, \textit{Embedding Model Evolution and Upgrades} is now routine.
To achieve better semantic representation and retrieval performance, embedding models are frequently updated through better architectures, larger training data, or higher embedding dimensions \cite{zhang2024wordvectorsmultimodalembeddings, lee2024nvembedimprovedtechniquestraining, lee2024geckoversatiletextembeddings, luo-etal-2024-large}. 
However, such upgrades often change the geometry, dimensionality, or semantics of the embedding space, making newly generated embeddings incompatible with existing ones.
As a result, organizations may need to operate multiple vectorDBs simultaneously rather than re-embedding and re-indexing the entire corpus at once.
For example, a system transitioning from a BERT-based model \cite{reimers-gurevych-2019-sentence, sanh2020distilbertdistilledversionbert} to a stronger GPT-based embedding model \cite{ni-etal-2022-large} may need to preserve access to both legacy and newly indexed data during the transition period.

Second, \textit{Cross-Domain Knowledge Retrieval} is also increasingly common~\cite{akkalyoncu-yilmaz-etal-2019-cross, nishida-etal-2023-sparse, melzer2023federated, guo-etal-2021-multireqa}. 
Many applications require retrieving information across multiple specialized domains, each with its own domain-specific embedding model~\cite{tang2024domain, 10.1145/3460824.3460826, chen2020domain, Ghosh, shen2022sscibert}. 
For example, a biomedical assistant may need to retrieve information jointly from genomics, pharmaceutical, and scientific literature databases, even when those corpora were embedded using different models optimized for different domains.


From a database systems perspective, this is a new federation problem.
Federated query processing has been extensively studied for structured databases~\cite{10.1145/96602.96604, 9773811, 9447055}, and recent work explores federation over vector databases~\cite{zhu2024fedsq, zhao2024frag}.
However, existing vector federation systems typically assume a shared embedding model and avoid the central challenge of heterogeneous settings: \textit{score comparability across incompatible embedding spaces}.
With different encoders, vectors cannot be directly compared and similarity scores are uncalibrated across databases.
Na"ively merging top-$K$ results therefore yields unstable global rankings and poor retrieval quality.
The key question is whether heterogeneous embeddings can be aligned to preserve retrieval quality while respecting real vectorDB construction and serving constraints.

A straightforward way to address this issue is to integrate all databases into a common embedding space by re-embedding the underlying corpus and rebuilding indexes using a unified model~\cite{Meng_2021_ICCV, Shen_2020_CVPR, seo2023metric, jaeckle2023fastfill, 10.1609/aaai.v37i3.25447}. While this can ensure compatibility, large-scale re-embedding and re-indexing incur substantial offline cost, system disruption, and migration complexity, especially for large or continuously updated corpora~\cite{drift-adapter, integration}. More importantly, heterogeneous vectorDB federation introduces a fundamental systems trade-off among three competing objectives: \textit{retrieval accuracy}, \textit{construction cost}, and \textit{query throughput}. Aggressive integration can improve retrieval quality but increases construction overhead, whereas lightweight approaches that avoid integration reduce upfront cost but often sacrifice ranking quality or runtime efficiency. A practical heterogeneous federation mechanism must therefore achieve score compatibility without paying the full cost of database unification.
 
To this end, in this paper, we propose \textbf{HVMap} (\textbf{H}eterogeneous \textbf{V}ector \textbf{Map}ping), a lightweight, system-agnostic compatibility layer that enables federated retrieval across heterogeneous vectorDBs without corpus migration or index reconstruction. 
It learns bidirectional mappings between embedding spaces to support cross-database score comparison, integrates with retrieval pipelines, and scales to multiple databases through a hub-based design. 
At its core, \textbf{HVMap} uses a retrieval-oriented fuzzy contrastive objective to align semantically related vectors across incompatible spaces while accommodating ambiguity across models and domains.

We evaluate \textbf{HVMap} in open-domain, cross-domain, and large-scale settings, and construct a benchmark of financial, biomedical, and scientific literature because existing benchmarks do not capture heterogeneous cross-domain federated retrieval.
Figure~\ref{fig:comparison} shows that \textbf{HVMap} occupies a favorable region of the quality--cost--throughput trade-off space.
Compared with existing alternatives, it achieves up to 3.59$\times$ higher nDCG@10, reduces construction time to 0.3\% of that required by baseline methods, and improves query throughput by up to 2.01$\times$.

Our contributions are summarized as follows.

\begin{itemize}[leftmargin=*, labelsep=0.5em]
\item We formulate federated retrieval across embedding-heterogeneous vectorDBs as a systems problem and identify cross-space score comparability as its central challenge.
\item We propose \textbf{HVMap}, a compatibility layer that preserves native vectors and indexes, combines bidirectional candidate transformation with global ranking, and scales to multiple vectorDBs through a bounded-depth hub topology with batched execution.
\item We construct and publicly release the first benchmark for cross-domain retrieval across heterogeneous vectorDBs, supported by human validation and independent LLM-based evaluation.
\item We demonstrate that \textbf{HVMap} achieves a favorable trade-off among retrieval accuracy, construction cost, and query throughput across open-domain, cross-domain, and large-scale settings.
\item We characterize the operating conditions of bidirectional mapping-based federation by analyzing embedding-space structural similarity, capacity, training objective, and score weighting.
\end{itemize}

\section{Motivation}

As vectorDBs evolve over time and across domains, federated retrieval over heterogeneous vectorDBs becomes increasingly common. Previous approaches such as compatible learning~\cite{Shen_2020_CVPR, hu2022learning, ramanujan2022forward, jaeckle2023fastfill, park2019relational} and domain alignment methods~\cite{wang-etal-2022-gpl, schopf-etal-2023-efficient, embeddingconverter, 10268909} have shown promise in representation alignment. 
However, they are not well suited to our target setting. 
Compatible learning typically assumes that a new embedding model is explicitly trained to remain compatible with an existing embedding space, while domain alignment methods usually require projecting stored data into a shared space. 
In our setting, where heterogeneous vectorDBs have already been constructed independently, these assumptions are often unrealistic or operationally expensive.

Recent work has explored more efficient integration techniques for heterogeneous vectorDBs~\cite{integration}, but this approach fundamentally assumes that the underlying databases can be integrated into a unified retrieval system. 
This assumption is often difficult to satisfy when vectorDBs are independently managed, deployed across separate systems, or specialized for different domains~\cite{q1_query, data_mesh, preivee}.

We therefore focus on a distinct systems problem: enabling federated retrieval across heterogeneous vectorDBs without corpus migration or index reconstruction. 
The goal is not to unify all databases into a single embedding space, but to enable comparable scoring across autonomous vectorDBs while balancing retrieval accuracy, construction cost, and query throughput.
This motivates \textbf{HVMap}, a lightweight compatibility layer that learns bidirectional mappings between embedding spaces and enables effective cross-database retrieval with modest construction and serving overhead.

\section{Methodology}
In this section, we formulate the problem and present \textbf{HVMap} together with existing approaches for federated retrieval across heterogeneous vectorDBs. 
For clarity, we begin with two vectorDBs and later extend the framework to more, as discussed in Section~\ref{sec:scalability}.

\subsection{Problem Formulation}
\label{sec:setup}

In a top-$K$ information retrieval system utilizing a vectorDB, the user expects to receive \( K \) most relevant documents for the submitted query. 
Let the vectorDB storing \(n = |D_\alpha^t|\) vectors be denoted as
\(D_\alpha^t = \{\mathbf{x}_{\alpha,1}^t, \mathbf{x}_{\alpha,2}^t, \dots, \mathbf{x}_{\alpha,n}^t\}\),
where \(t\) indicates the corresponding embedding space and \(\alpha\) identifies the document set stored in this vectorDB.
In other words, the vectorDB \( D_\alpha^t \) is constructed using the embedding model \( \phi^t : \text{text} \to \mathbb{R}^{dim_t} \), which maps text data into the embedding space \( t \) with dimension \( dim_t \).
Each element stored in \( D_\alpha^t \), \( \mathbf{x}_{\alpha,i}^t \), where \( i \in \{1,2,\dots,n\} \), represents a vector that retains the semantics of its corresponding original document in the set \( \alpha \).
Then, when a query is submitted to the system, it must first be converted into vector form. 
Let the query in text form be denoted as \( text_q \). Then, the vector representation of the given query can be obtained as \( \mathbf{q}^t = \phi^t(text_q) \).
Now, we perform a vector search to retrieve the top \( K \) relevant vectors from \( D_\alpha^t \), which can be expressed as:
\[
\qquad \text{SEARCH} (D_\alpha^t, K, \mathbf{q}^t) = \underset{\mathbf{x} \in D_\alpha^t}{\arg\text{Top}K} \, sim(\mathbf{q}^t, \mathbf{x})
\]
where \( sim \) represents the similarity function that quantifies the closeness between two vectors, and \( \arg\text{Top}K \) returns the top \( K \) elements from the given vector set that maximize the given similarity function. 
The vectors are sorted in descending order of their function values; here, the ordering is based on similarity to the query vector.
A larger value of the function $sim$ indicates the two vectors are more similar. 
Note that the SEARCH operation is only possible when $D$ and $\textbf{q}$ from \( \arg\text{Top}K \) share the same superscript (i.e., belong to the same embedding space). 
Otherwise, the query vector is not compatible with the vectors in \( D \), which can significantly degrade performance or even cause system failure. 

Extending the formulation to federated queries over multiple vectorDBs, the retrieval across two heterogeneous vectorDBs, \( D_\alpha^t \) and \( D_\beta^s \), can be expressed as follows:
\[
\text{SEARCH}(D_\alpha^t, D_\beta^s, K, \mathbf{q}^t) = \underset{\mathbf{x} \in D_\alpha^t \cup D_\beta^s}{\arg\text{Top}K} \,  sim( \mathbf{q}^t, \mathbf{x}).
\]
This SEARCH operation is expected to retrieve a total of \( K \) relevant vectors from the two vectorDBs.
At this point, since the vectors in \( D_\beta^s \) and the query vector \( \mathbf{q}^t \) reside in different embedding spaces, a compatibility issue arises, meaning that their similarity cannot be properly computed using the similarity function \( sim \).
In the worst case, this incompatibility results in complete system failure when the embedding dimensions differ, as similarity cannot be computed across vectors from different spaces. 
Even when the dimensions coincide, vectors from \(D_\beta^s\) may still be assigned low similarity scores despite their true semantic relevance to the query. 
Consequently, they are unlikely to appear in the top \(K\) results, which prevents effective retrieval from \(D_\beta^s\) and degrades overall retrieval quality. 
Addressing this compatibility issue is therefore a fundamental requirement for federated retrieval across heterogeneous vectorDBs.

\subsection{Conventional Approaches}
\label{sec:baselines}

Conventional approaches to heterogeneous vectorDB federation broadly fall into migration-based integration and result fusion.
\textbf{Migration (MIG)}~\cite{Meng_2021_ICCV,Shen_2020_CVPR} resolves embedding incompatibility by re-embedding the documents of one vectorDB with the embedding model of the other and rebuilding a unified index. 
When migrating \(D_\beta^s\) into space \(t\), the resulting database is
$
D_{\alpha\cup\beta}^t
=
D_\alpha^t
\cup
\{\phi^t(\text{doc}_{\beta,i})\}_{i=1}^{n},
$
and retrieval is performed as
\[
\text{SEARCH}(D_\alpha^t,D_\beta^s,K,\mathbf{q}^t)
=
\text{SEARCH}(D_{\alpha\cup\beta}^t,K,\mathbf{q}^t).
\]
\textbf{MIG} ensures score compatibility, but requires corpus-wide re-embedding and index reconstruction.

Fusion-based methods instead query each vectorDB using its corresponding embedding model and combine the returned ranked lists. 
Let \(A=\langle\mathbf{a}_1,\ldots,\mathbf{a}_K\rangle\) and
\(B=\langle\mathbf{b}_1,\ldots,\mathbf{b}_K\rangle\) denote the two local result lists. 
\textbf{Aggregation} (\textbf{AGG})~\cite{AGG} alternates candidates from \(A\) and \(B\):
\[
\text{AGG}(A,B)
=
\langle
\mathbf{a}_1,\mathbf{b}_1,
\mathbf{a}_2,\mathbf{b}_2,
\ldots
\rangle_{[:K]},
\]
where \((\cdot)_{[:K]}\) denotes truncation to the first \(K\) candidates.
AGG implicitly assumes that relevant documents are approximately balanced across sources, and its ranking quality can therefore degrade under source-skewed relevance.

\textbf{Reciprocal Rank Fusion (RRF)}~\cite{rrf} assigns each candidate a score based on its rank across the source lists:
\[
\text{RRF}(\mathbf{x})
=
\sum_{\ell\in\{A,B\}}
\frac{1}{c+\operatorname{rank}_{\ell}(\mathbf{x})},
\]
where \(c\) is a smoothing constant.
\textbf{CombMNZ}~\cite{combmnz} instead fuses normalized local similarity scores:
\[
\text{CombMNZ}(\mathbf{x})
=
N(\mathbf{x})
\sum_{\ell\in\{A,B\}}
\widehat{s}_{\ell}(\mathbf{x}),
\]
where \(\widehat{s}_{\ell}(\mathbf{x})\) is the normalized local similarity score of candidate \(\mathbf{x}\) within source \(\ell\), and \(N(\mathbf{x})\) is the number of source lists containing \(\mathbf{x}\).
Both methods globally rank the \(2K\) candidates and return the final top-\(K\) results.
These fusion methods avoid migration and index reconstruction, but require one query embedding per vectorDB, introducing additional query-time overhead, and do not explicitly align heterogeneous embedding spaces.
Accordingly, \textbf{MIG}, \textbf{AGG}, \textbf{RRF}, and \textbf{CombMNZ} serve as representative conventional baselines, while Section~\ref{sec:expr} further considers learning-based alternatives.

\begin{figure}[t!]
  \centering
  \includegraphics[width=0.9\linewidth]{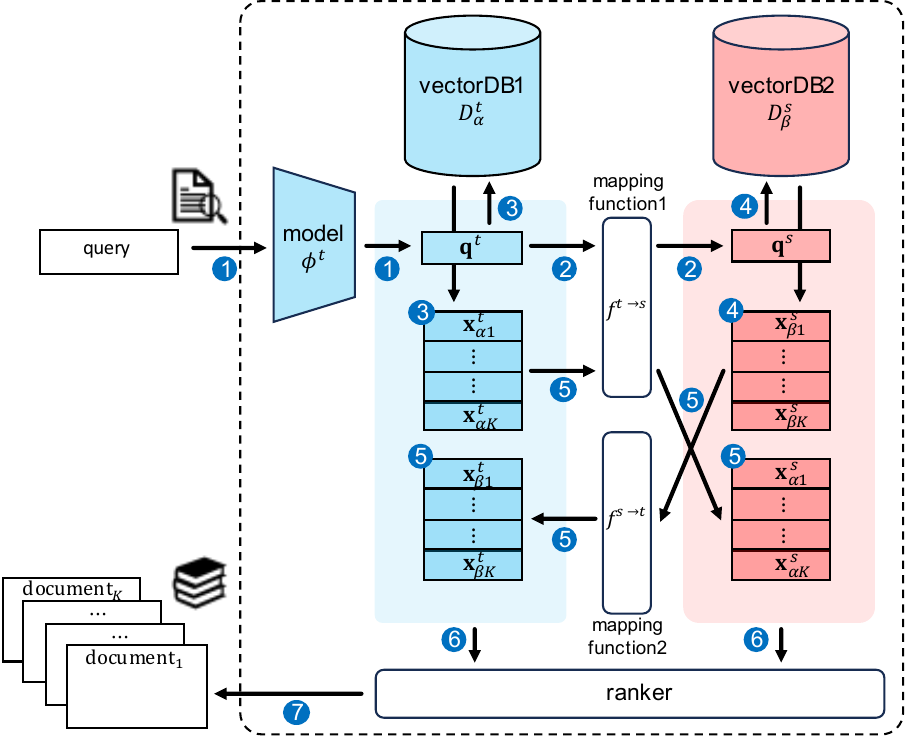}
  \caption{A framework of \textbf{HVMap} for federated retrieval across two heterogeneous vectorDBs.}
  \label{fig:framework}
\end{figure}

\subsection{Overview of HVMap}
\label{sec:pipeline}

Existing approaches struggle to balance construction cost, retrieval accuracy, and query throughput in federated retrieval.
\textbf{MIG} preserves retrieval quality by unifying heterogeneous vectorDBs into a common space, but requires costly re-embedding and index reconstruction.
In contrast, fusion-based federations avoid these construction costs by querying each vectorDB independently, but suffer from limited score comparability and the overhead of running multiple embedding models at query time.
\textbf{HVMap} addresses this trade-off through a lightweight compatibility layer based on bidirectional embedding-space mappings learned with fuzzy contrastive loss, enabling heterogeneous vectorDBs to remain physically separate while being exposed as a single logical retrieval system.

As shown in Figure~\ref{fig:framework}, the blue and red colors represent embedding spaces $t$ and $s$, respectively.
Vectors from different spaces are not directly compatible, even when they encode semantically related documents.
When a query is submitted, it is first encoded by the predefined embedding model $\phi^t$ of one vectorDB:
$
\mathbf{q}^t = \phi^t(\textit{text}_q)
$
(Figure~\ref{fig:framework}.\filledcircled{1}).
To search the second vectorDB without invoking another full embedding model, \textbf{HVMap} transforms the query into space $s$ using the learned mapping function
$f^{t\rightarrow s}:\mathbb{R}^{dim_t}\rightarrow\mathbb{R}^{dim_s}$:
$
\mathbf{q}^s \approx f^{t\rightarrow s}(\mathbf{q}^t)
$
(Figure~\ref{fig:framework}.\filledcircled{2}).
Once query vectors are available in both spaces, \textbf{HVMap} independently searches the two vectorDBs, $D_\alpha^t$ and $D_\beta^s$, yielding
\[
\mathbf{X}_\alpha^t
\coloneqq
\text{SEARCH}(D_\alpha^t,K,\mathbf{q}^t)
,\; \text{and} \;
\mathbf{X}_\beta^s
\coloneqq
\text{SEARCH}(D_\beta^s,K,f^{t\rightarrow s}(\mathbf{q}^t))
\]
(Figure~\ref{fig:framework}.\filledcircled{3}--\filledcircled{4}).
Because the mapping function is substantially lighter than a full embedding model, this design reduces query-time encoding overhead compared with methods that independently encode the query using every participating model.
From the user's perspective, \textbf{HVMap} exposes the heterogeneous vectorDBs as a single logical retrieval system, even though retrieval is performed independently in their original embedding spaces.

At this stage, the two vectorDBs have returned $2K$ candidates,
$\mathbf{X}_\alpha^t \cup \mathbf{X}_\beta^s$.
However, their local similarity scores cannot be directly compared because they originate from different embedding spaces and may follow different score distributions.
This makes simple score merging unreliable.
To address this issue, \textbf{HVMap} maps each retrieved candidate into the opposite embedding space using
$f^{t\rightarrow s}$ and $f^{s\rightarrow t}$
(Figure~\ref{fig:framework}.\filledcircled{5}).
The ranker can therefore evaluate every candidate using two complementary signals: its similarity in the native space, where no mapping error is introduced, and its similarity in the mapped space, which provides cross-space evidence for global calibration
(Figure~\ref{fig:framework}.\filledcircled{6}).

For a candidate $\mathbf{x}^k$ originating from embedding space
$k\in\{t,s\}$, the final ranking score is computed as
\[
\text{SCORE}_2(\mathbf{x}^k)=
\begin{cases}
w_1 \cdot sim(\mathbf{q}^t,\mathbf{x}^k)
+
w_2 \cdot sim(\mathbf{q}^s,f^{t\rightarrow s}(\mathbf{x}^k)),
& \text{if } k=t, \\[2mm]
w_1 \cdot sim(\mathbf{q}^s,\mathbf{x}^k)
+
w_2 \cdot sim(\mathbf{q}^t,f^{s\rightarrow t}(\mathbf{x}^k)),
& \text{if } k=s.
\end{cases}
\]
Following the notation in Section~\ref{sec:setup},
$\mathbf{x}^k$ denotes a vector originally constructed in embedding space $k$.
The weights satisfy $w_1,w_2\geq 0$ and $w_1+w_2=1$, where $w_1$ controls the native-space contribution and $w_2$ controls the mapped-space contribution.
The analysis in Section~\ref{sec:weight_study} shows that these two signals are complementary but not equally reliable.
Assigning slightly greater weight to the native-space similarity improves retrieval performance by using the error-free native score as the primary signal, while retaining the mapped-space similarity as a complementary signal for comparison.

After computing $\text{SCORE}_2$ for all retrieved candidates, the ranker globally selects the $K$ highest-scoring vectors and returns their corresponding documents
(Figure~\ref{fig:framework}.\filledcircled{7}):
\[
\text{SEARCH}(D_\alpha^t,D_\beta^s,K,\mathbf{q}^t)
=
\underset{
\mathbf{x}\in
\mathbf{X}_\alpha^t\cup\mathbf{X}_\beta^s
}{
\arg\text{Top} \:
}
\text{SCORE}_2(\mathbf{x}).
\]
Through this pipeline, \textbf{HVMap} enables score-compatible federated retrieval without modifying the stored vectors or rebuilding the underlying indexes.

\subsubsection{Budget under Skewed Relevance}
Determining which sources to search and how many candidates to retrieve from each has long been studied as a resource-selection problem in federated retrieval~\cite{si2002language,si2003redde}.
In particular, this problem becomes important when relevant documents are unevenly distributed across sources.
\textbf{HVMap} handles such skew without imposing a fixed final quota on each vectorDB, while allowing every source to contribute up to all $K$ final results.
Given a requested output size $K$, it retrieves up to $K$ candidates from every participating vectorDB and globally reranks the resulting candidate pool using $\text{SCORE}_2$.
Each source therefore has an equal opportunity to contribute all $K$ final results, and the final top-$K$ documents may originate entirely from a single vectorDB even when relevance is highly skewed.
Unlike interleaving-based aggregation, \textbf{HVMap} does not assume that relevant documents are evenly distributed across sources; Section~\ref{sec:skewed_relevance} evaluates this property empirically.
However, \textbf{HVMap} inherits the candidate-generation limitation of retrieve-and-rerank systems: documents excluded from a source's local top-$K$ cannot be recovered globally.
Retrieving $K$ candidates per source is a conservative default that does not restrict any source's final contribution.
For larger federations, adaptive resource selection could reduce retrieval and mapping overhead by allocating smaller budgets to unlikely sources.
We leave this extension to future work.

\begin{figure}[t!]
  \centering
  \includegraphics[width=0.9\linewidth]{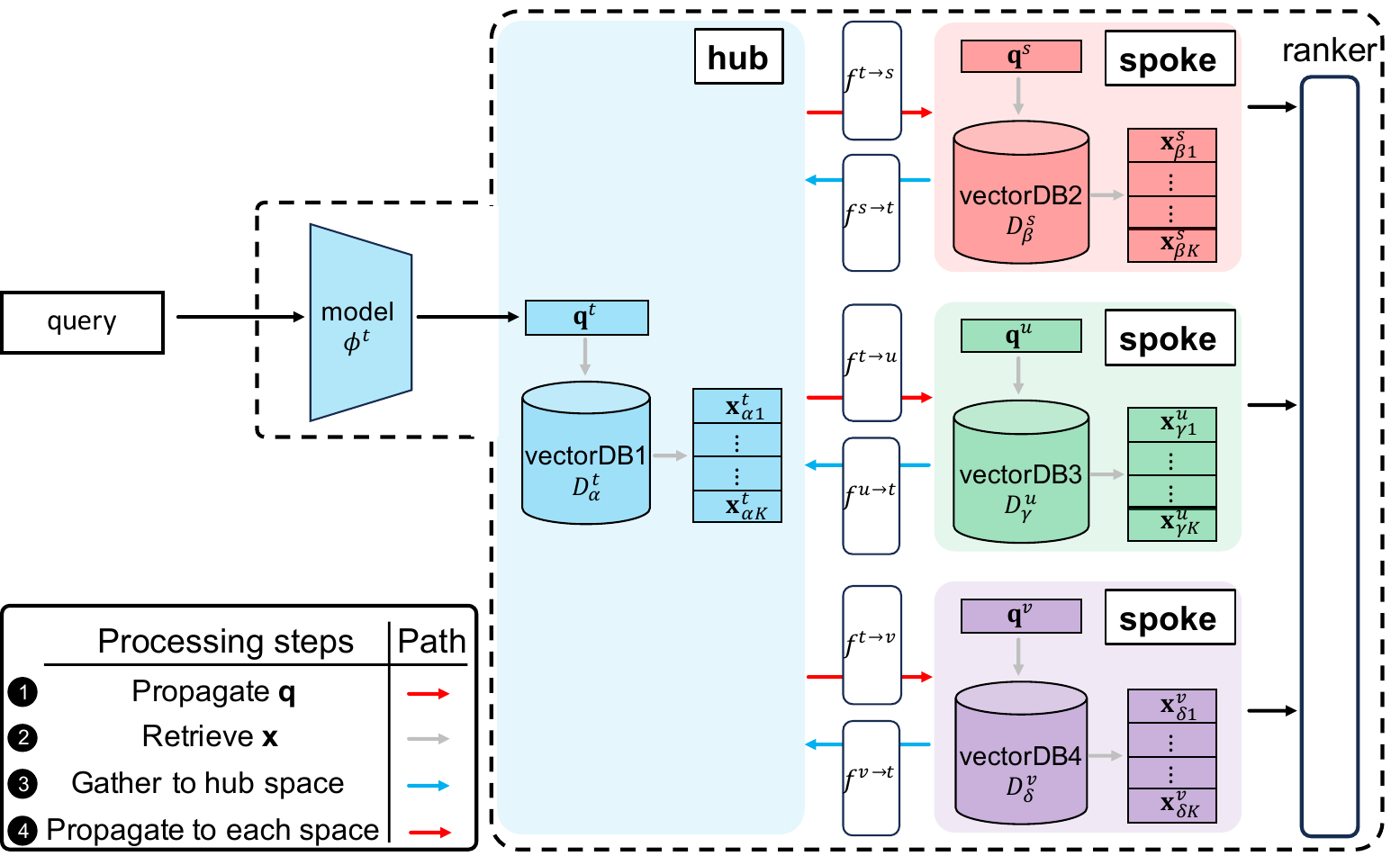}
    \caption{Extended \textbf{HVMap} framework for scalable federated retrieval across four heterogeneous vectorDBs. 
    }
  \label{fig:scalability}
\end{figure}

\subsection{Scalability of HVMap}
\label{sec:scalability}
Up to Section~\ref{sec:pipeline}, we have focused on federated retrieval across two heterogeneous vectorDBs. 
In this section, we extend our discussion to illustrate how \textbf{HVMap} scales to an arbitrary number of vectorDBs. 
As an example, we consider a system managing four vectorDBs, 
$D_\alpha^t$, $D_\beta^s$, $D_\gamma^u$, and $D_\delta^v$, as shown in Figure~\ref{fig:scalability}.

\subsubsection{Process of Federated Retrieval at Scale}
The fundamental process of mapping the query and retrieved vectors to each embedding space remains the same as described in Section~\ref{sec:pipeline}. 
However, scaling to multiple vectorDBs introduces two additional considerations: the increased number of embedding spaces, which leads to a larger number of mapping functions, and the connectivity structure among them.
To enable scalable retrieval across many vectorDBs, \textbf{HVMap} adopts a \textit{hub-and-spoke} architecture, a widely used design in distributed systems and network architectures for reducing communication overhead and simplifying system connectivity~\cite{hub1, hub2}. 
In this design, one embedding space acts as the hub, while the remaining spaces serve as spokes.
When a query requiring retrieval of the top-$K$ relevant vectors arrives, it is first encoded by the hub embedding model \(\phi^t\), yielding the query vector \(\mathbf{q}^t\).
The query vector is then propagated to each spoke embedding space via the corresponding mapping functions, following the red-arrow paths (Figure~\ref{fig:scalability}.\filledcircled{1}), to produce query representations in all embedding spaces.
Once the query vectors are available in every space, each of the four vectorDBs independently retrieves $K$ candidate vectors (Figure~\ref{fig:scalability}.\filledcircled{2}), yielding $4K$ candidates in total.

To perform ranking, these vectors must be made comparable across embedding spaces. 
HVMap performs this through a two-stage gather-and-propagate process. 
First, the retrieved vectors from each spoke are mapped back to the hub space (blue arrows in Figure~\ref{fig:scalability}.\filledcircled{3}), where they are gathered into a unified candidate pool. 
The entire candidate pool is then propagated from the hub to each spoke space in a single batch, excluding the vectors originally retrieved from that spoke (red arrows in Figure~\ref{fig:scalability}.\filledcircled{4}).
This ensures that every embedding space receives candidate vectors originating from all other spaces.
Finally, the vectors and their corresponding queries are passed to the ranker, which extends $\text{SCORE}_2$ to select the final top-$K$ documents. 
A fixed weighting provides a practical default, while federation-specific tuning may further improve performance; see Section~\ref{sec:lim_dis}.

\subsubsection{Advantages of the Hub-Based Architecture}
The hub-and-spoke design improves scalable federated retrieval in construction cost, accuracy, and throughput.

\noindent\textbf{Construction Cost.}
From the perspective of construction cost, the hub architecture significantly reduces the number of required mapping functions.
To guarantee vector compatibility, transformations must be supported between any pair of embedding spaces and the most intuitive approach is to maintain pairwise connections among $N$ embedding spaces, which requires $O(N^2)$ mapping functions, where $N$ denotes the number of vectorDBs.
In contrast, the hub-based design composes transformations between two arbitrary embedding spaces $s$ and $r$ via the hub.
As a result, only $O(N)$ mappings are required, keeping the construction overhead manageable even as the number of vectorDBs increases.
Moreover, choosing a lightweight embedding model as the hub can further reduce mapping function training time, since the hidden dimension of each mapping function is determined solely by the input and output embedding dimensions.

\noindent\textbf{Retrieval Accuracy.}
The hub-based architecture also bounds the propagation of mapping errors, since each mapping step may introduce a small amount of distortion. 
Any cross-space transformation requires at most two steps (spoke \(\rightarrow\) hub \(\rightarrow\) spoke), limiting the accumulation of such errors. 
By contrast, a linear chain of embedding spaces could lead to a mapping depth of up to \(N-1\). 
Because transformations between non-hub spaces are mediated through the hub, the choice of hub embedding model can influence the final retrieval accuracy. 
A stronger hub model may improve alignment quality, although at the cost of higher query embedding latency. 
This design therefore allows system designers to select the hub model based on their preferred trade-off between accuracy and query efficiency.\footnote{In our experiments, the choice of hub embedding model led to only minor accuracy differences, whereas using a lightweight model as the hub noticeably improved query throughput. We therefore recommend the latter in practice.}

\noindent\textbf{Query Throughput.}
The hub-based architecture also improves query throughput in two respects. 
First, it allows a lightweight embedding model to be used at the hub, reducing the cost of query embedding. 
Second, it supports an efficient two-stage gather-and-propagate mechanism for cross-space vector transformation. 
Rather than sequentially exchanging vectors between every pair of spaces, HVMap first gathers retrieved vectors into the hub space and then redistributes them to the spokes. 
This design enables batch mapping of the full set of \(NK\) candidate vectors, thereby improving computational efficiency. More detailed empirical measurements and quantitative analysis are presented in Section~\ref{sec:scale_expr}.

\subsection{Embedding Space Mapping Function}
The mapping function $f$ is a core component of \textbf{HVMap}, enabling compatibility between heterogeneous spaces.
Trained with the proposed fuzzy contrastive loss, it learns a transformation such that
$f(\phi^t(\text{doc})) \approx \phi^s(\text{doc})$.
Given a document set
$\mathcal{T}=\{\text{doc}_1,\dots,\text{doc}_N\}$,
we construct paired embeddings
$\mathbf{X}^t=\phi^t(\mathcal{T})=\{\mathbf{x}_1^t,\dots,\mathbf{x}_N^t\}$
and
$\mathbf{X}^s=\phi^s(\mathcal{T})=\{\mathbf{x}_1^s,\dots,\mathbf{x}_N^s\}$.
Since $\mathbf{x}_i^t$ and $\mathbf{x}_i^s$ are generated from the same $\text{doc}_i$, they represent the same semantics.

\subsubsection{Fuzzy Contrastive Loss}
\label{sec:fuzzy_loss}

Since our mapping task can be viewed as supervised learning over paired embeddings from two spaces, a natural baseline is to minimize the mean squared error (MSE) between mapped embeddings and their paired targets.
This approach, also adopted in prior work~\cite{jaeckle2023fastfill,ramanujan2022forward,drift-adapter}, provides a simple solution for embedding transformation.
However, pointwise regression alone is often insufficient for learning high-quality cross-space alignment~\cite{embeddingconverter}.
Contrastive learning, as used in CLIP~\cite{pmlr-v139-radford21a}, additionally encourages positive pairs to be close and negative pairs to be separated.
Its basic form is
\[
\hspace*{-0.5em}
\mathcal{L}_{\mathrm{CL}}
=
y_{ij}\left\|f(\mathbf{x}_i^t)-\mathbf{x}_j^s\right\|^2
+
(1-y_{ij})
\max\left(
0,
m-\left\|f(\mathbf{x}_i^t)-\mathbf{x}_j^s\right\|
\right)^2,
\]
where $y_{ij}=1$ for a positive pair, i.e., $i=j$, and $y_{ij}=0$ otherwise.
The first term aligns positive pairs, while the second penalizes negative pairs whose distance is smaller than the fixed margin $m$.

Although standard contrastive loss (CL) is effective in many settings, it treats all negative pairs using the same margin regardless of their semantic dissimilarity.
In practice, negative pairs may differ substantially in how far apart they should be.
Motivated by this observation, we generalize the fixed margin into an adaptive, pair-dependent margin $\mu_{ij}$:
\begin{align*}
\mathcal{L}_{\mathrm{fuzzy}}
=
y_{ij}&\left\|f(\mathbf{x}_i^t)-\mathbf{x}_j^s\right\|^2 \\
&+
(1-y_{ij})
\left[
\max\left(
0,
\mu_{ij}-\left\|f(\mathbf{x}_i^t)-\mathbf{x}_j^s\right\|
\right)
\right]^2.
\end{align*}
Here, $\mu_{ij}$ encodes the semantic dissimilarity of each pair.
Semantically distant pairs receive larger margins, while closely related pairs receive smaller margins and therefore weaker separation.
Because $\mu_{ij}$ varies continuously across pairs, the loss provides finer-grained supervision than a uniform margin.

\begin{figure}[t!]
  \centering
  \includegraphics[width=0.9\linewidth]{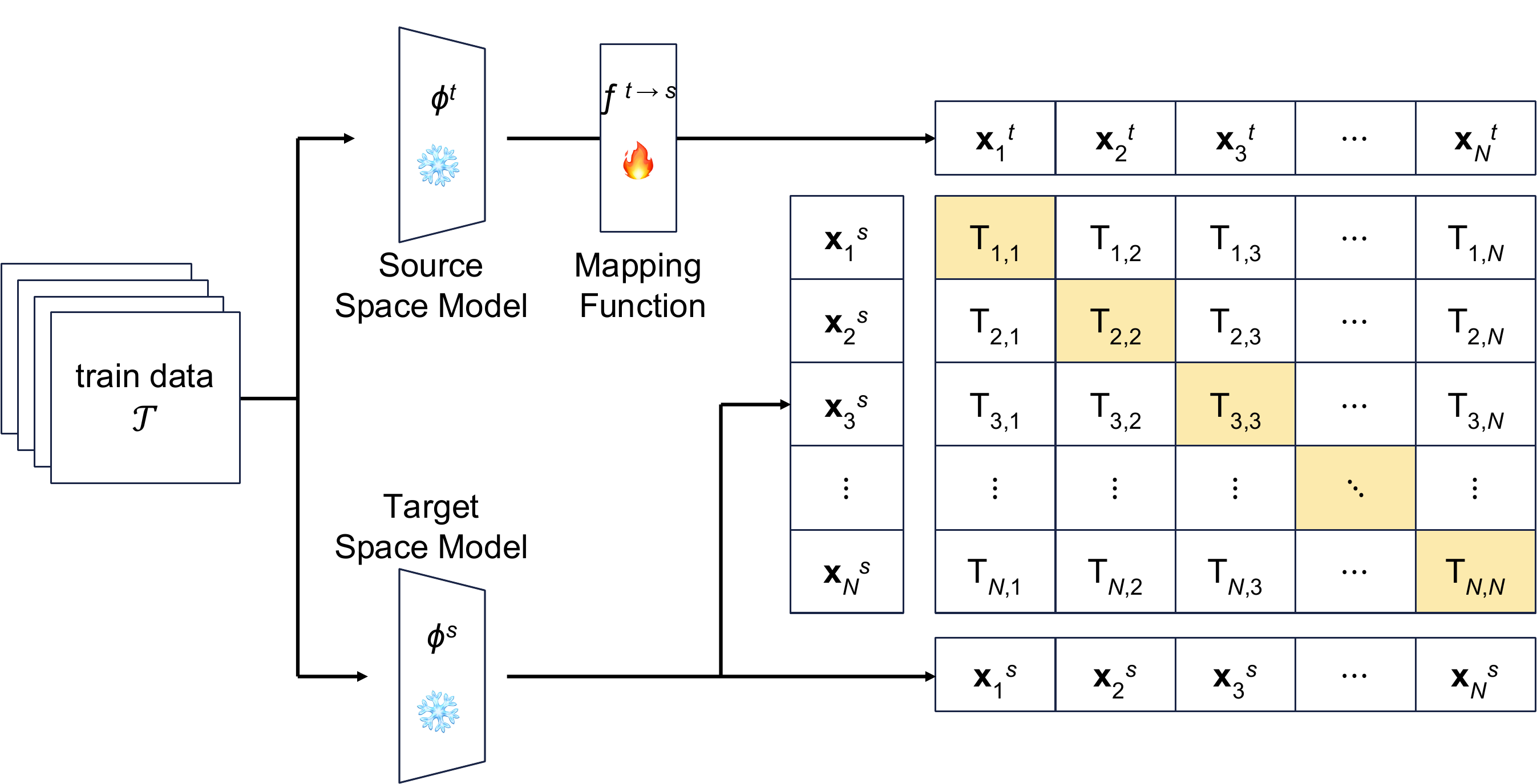}
  \caption{Visualization of the fuzzy contrastive loss used to train the mapping function. The sum of $T_{i,j}$ constitutes the final loss, where the highlighted cells represent positive pairs, and the remaining cells correspond to negative pairs.}
  \label{fig:fuzzy_loss}
\end{figure}

We instantiate this idea by defining $\mu(\mathbf{x}_i^s,\mathbf{x}_j^s)$ as the semantic distance between $\mathrm{doc}_i$ and $\mathrm{doc}_j$, thereby leveraging the full pairwise structure of the target embedding space $s$.
When the two target embeddings are similar, $\mu$ approaches zero; otherwise, it increases with their dissimilarity.
Since $\mu(\mathbf{x}_i^s,\mathbf{x}_i^s)=0$, positive pairs become a special case of the same pairwise relation.
Removing the hinge clipping and directly regressing each mapped-to-target distance toward this target-space relation yields
\[
\mathcal{L}(\mathbf{X}^t,\mathbf{X}^s;f)
=
\sum_{i=1}^{N}\sum_{j=1}^{N}
\left\|
\mu(\mathbf{x}_i^s,\mathbf{x}_j^s)
-
d\!\left(f(\mathbf{x}_i^t),\mathbf{x}_j^s\right)
\right\|^2,
\]
where $d(\cdot,\cdot)$ denotes a distance measure.
As illustrated in Figure~\ref{fig:fuzzy_loss}, with both embedding models frozen, the diagonal terms $T_{i,i}$ align each mapped embedding with its paired target, while the off-diagonal terms preserve its graded relations to the remaining target vectors.
This matrix-form objective therefore unifies pointwise alignment and contrastive relational supervision within a single formulation that is well suited to efficient batch computation.

\subsubsection{Relation to Relational Knowledge Distillation}
\label{sec:rkd_relation}

Although Fuzzy CL and Relational Knowledge Distillation (RKD) ~\cite{park2019relational} are motivated by different objectives, both exploit pairwise relational structure.
In our setting, RKD objective can be written as
\[
\mathcal{L}_{\mathrm{RKD}}
=
\sum_{i=1}^{N}\sum_{j=1}^{N}
\left\|
d\!\left(\mathbf{x}_i^s,\mathbf{x}_j^s\right)
-
d\!\left(
f(\mathbf{x}_i^t),
f(\mathbf{x}_j^t)
\right)
\right\|^2.
\]
RKD preserves target-space geometry by matching \emph{mapped-to-mapped} distances to the corresponding \emph{target-to-target} distances.
Fuzzy CL differs in the relation it optimizes. 
Rather than matching the internal geometry among mapped embeddings, it matches \emph{mapped-to-target} distances to the semantic relations defined in the target space.
This formulation is aligned with the actual retrieval operation: the target vectorDB remains indexed by the original vectors
$\mathbf{x}_j^s$, and a mapped query is directly compared with these stored target vectors.
Fuzzy CL therefore preserves the score profile consumed by retrieval over the unchanged target index.
This structural difference also provides explicit pointwise anchoring.
For $i=j$, Fuzzy CL enforces $d(f(\mathbf{x}_i^t),\mathbf{x}_i^s)\approx 0$, directly aligning each mapped embedding with its paired target.
In contrast, the corresponding RKD term compares an embedding with itself and is therefore always zero, so RKD alone cannot determine the absolute location of mapped embeddings in the target space.
The empirical differences between Fuzzy CL and RKD are analyzed in Section~\ref{sec:loss_compare}.

\subsubsection{Implementation}
\label{sec:implementation}
The mapping function is implemented as an MLP with SELU activation and a normalization layer, following prior work~\cite{jaeckle2023fastfill,ramanujan2022forward,embeddingconverter}.
Its depth and hidden dimension are treated as hyperparameters and set to four layers and five times the output dimension, respectively, with their impact analyzed in Section~\ref{sec:mapping_ablation}.
For the function $\mu$, since all target data are text embeddings, we define $\mu$ using cosine distance:
\[
\mu(\mathbf{x}_1,\mathbf{x}_2) = 1 -
\frac{\mathbf{x}_1 \cdot \mathbf{x}_2}{\|\mathbf{x}_1\| \cdot \|\mathbf{x}_2\|}.
\]
The distance function $d$ follows the metric used in the corresponding embedding space. 
Note that $\mu$ is flexible and can be adapted to other similarity functions, including manually designed measures.
Ideally, the loss is computed over all training pairs. 
However, for computational feasibility, we use mini-batch training with a batch size of 1792, randomly reshuffled at each epoch. 
The model is trained for 40 epochs with a learning rate of 0.001.

\section{Cross-Domain Retrieval Benchmark}
\label{sec:dataset_generation}

To evaluate our method in cross-domain settings, we require a benchmark consisting of \textit{query--relevant corpus} pairs, where the relevant corpora span multiple domains and must be jointly retrieved to answer a single query. An illustrative example is shown below:
\begin{tcolorbox}[colback=gray!10, colframe=black, width=\linewidth]

\textbf{- Query: }
How could advancements in gene-editing and AI diagnostic systems contribute during a financial crisis?

\textbf{- Corpus 1 (Financial): }
The global financial crisis had a severe impact on the economies of many countries. \ldots

\textbf{- Corpus 2 (Biomedical): }
Modern medicine is incorporating innovative technologies in diagnosing and treating various diseases. \ldots
\end{tcolorbox}


\noindent However, no publicly available benchmark requires retrieval across multiple domain-specific corpora for a single query. 
We therefore construct a cross-domain retrieval benchmark using an LLM-based pipeline that generates candidate queries and refines them through multi-stage validation. 
Prior studies and open-source frameworks demonstrate the feasibility of LLM-generated IR benchmarks and systematic query refinement~\cite{SIGIR2024, dai-etal-2024-cocktail, ragas_doc}.

\subsection{Cross-Domain Query Generation Using LLM}
The source corpora comprise independently curated datasets from distinct knowledge domains, selected so that the generated queries require retrieval across domains.
For each sample, one corpus from each domain-specific dataset is randomly selected and provided to an LLM, which generates a query requiring information from all selected corpora.
This design ensures that no individual corpus alone is sufficient to fully answer the query.
The benchmark is expanded by varying domain combinations to create cross-domain evaluation sets.
Using OpenAI's GPT-4o~\cite{openai2024gpt4osystemcard}, we generate 500 queries for each benchmark set.
The resulting datasets cover four domain combinations: Financial--Science, Financial--Biomedical, Biomedical--Science, and Financial--Science--Biomedical.

\subsection{Validation of Relevant Corpus Set}
The corpora used to generate a query are not necessarily its most semantically relevant sources. 
We therefore refine each relevant corpus set through vector-based retrieval. 
For each domain, we construct a Chroma vectorDB~\cite{chromadb} using OpenAI's text-embedding-3-small model~\cite{openai2024embedding}, retrieve additional candidate corpora for each generated query, and add a retrieved corpus when it is more relevant than an initially selected one.
\rb{Because corpus-set expansion uses a single commercial embedding model, relevant corpora outside its retrieval pool may remain unjudged, potentially disadvantaging methods with different retrieval behavior.}

\ra{\textit{Human validation.}
To assess the consistency of the resulting labels, we randomly sampled 100 queries from each of the four cross-domain settings, yielding 400 queries and 2,564 associated query--document pairs. 
Three authors independently made binary relevance judgments for every pair. 
They labeled 99.0\%, 98.8\%, and 98.3\% of the pairs as relevant, respectively, with a mean pairwise agreement of 99.1\% and Fleiss' $\kappa=0.64$~\cite{fleiss1971measuring}. 
The moderate chance-corrected agreement reflects the highly imbalanced labels, as nearly all sampled pairs were judged relevant. 
Table~\ref{tab:benchmark_validation} reports the results.}\footnote{The cross-domain benchmark is available at \url{https://huggingface.co/datasets/snu-aidas/HVMap-dataset}.}

\begin{table}[t]
\centering
\footnotesize
\caption{\ra{Human validation of the cross-domain benchmark.}}
\label{tab:benchmark_validation}
\begin{tabular}{lccc}
\toprule
Annotator & \# Queries & \# Pairs & Relevant ratio \\
\midrule
Annotator 1 & 400 & 2,564 & 99.0\% \\
Annotator 2 & 400 & 2,564 & 98.8\% \\
Annotator 3 & 400 & 2,564 & 98.3\% \\
\midrule
\multicolumn{4}{c}{Mean pairwise agreement: 99.1\%; Fleiss' $\kappa$: 0.64} \\
\bottomrule
\end{tabular}
\end{table}

\section{Experiments}
\label{sec:expr}
In this section, we evaluate \textbf{HVMap} against multiple baselines in open-domain and cross-domain settings, focusing on retrieval quality, scalability, construction cost, and query throughput.

\subsection{Experimental Setup}
\subsubsection{Dataset}
The benchmark datasets are categorized into open-domain and domain-specific collections. 
For open-domain experiments, we use six datasets from BEIR~\cite{thakur2021beir}: Msmarco, Nq, Hotpotqa, Arguana, Dbpedia-Entity, and FEVER. 
For cross-domain experiments, we consider financial, biomedical, and scientific domains using FIQA~\cite{maia2018www}, BioASQ~\cite{bioasq}, and a combination of SciDocs~\cite{cohan2020specter} and SciFact~\cite{scifact}, respectively.
For each federated retrieval experiment, the documents are randomly partitioned into equally sized, disjoint subsets, with the number of subsets matching the number of target vectorDBs. 
In the cross-domain setting, each domain forms a dedicated vectorDB and is embedded using its corresponding domain-specific embedding model.
For \textbf{HVMap}, the mapping functions were trained using five of the open-domain BEIR datasets, excluding ArguAna. 
We randomly sampled 500,000 examples from each dataset, resulting in 2.5 million training samples.


\subsubsection{Embedding Model Selection}
For the open-domain setting, we used three Sentence-Transformer models: MPNet~\cite{NEURIPS2020_c3a690be}, DistilBERT~\cite{sanh2020distilbertdistilledversionbert}, and MiniLM~\cite{NEURIPS2020_3f5ee243}. 
For scalability experiments, we further included DistilRoBERTa~\cite{roberta}, Sentence-T5~\cite{t5}, and LaBSE~\cite{LaBSE}.
For domain-specific models, we leveraged specialized Sentence-Transformer models tailored for each domain. 
Specifically, the financial model (\textit{Fin}) used was \textit{financial-rag-matryoshka}\cite{financial-rag-matryoshka}, the biomedical model (\textit{Bio}) was \textit{bge-base-bioasq-matryoshka}\cite{bge-base-bioasq-matryoshka}, and the science model (\textit{Sci}) was \textit{S-PubMedBERT}\cite{deka2022improved}.
\ra{Our evaluation focuses on text because \textbf{HVMap} is motivated by RAG-style retrieval over independently constructed text-embedding vectorDBs.}

\subsubsection{Hyperparameter Configuration}
\label{sec:hyperparameter}
Unless otherwise specified, all vectorDBs were constructed using the HNSW (Hierarchical Navigable Small World) index~\cite{HNSW} implemented in Faiss~\cite{faiss2024}, with $M=32$, $\textit{efC}=256$, and $\textit{efS}=256$, while the retrieval cutoff was set to $K=10$.
For the weighted $\text{SCORE}_2$ function, we use a single global setting of $w_1=0.7$ across all datasets and embedding-model pairs, without per-dataset or per-pair tuning.
Section~\ref{sec:weight_study} analyzes the sensitivity of retrieval performance to this weight and shows that performance remains stable over a broad range of values.
For scalability experiments, we assign equal weight to the similarity contribution from each participating embedding space.
The effect of HNSW hyperparameters is analyzed in Section~\ref{sec:tuned}.


\subsubsection{Baseline Methods}
\label{sec:method_config}

We compare \textbf{HVMap} with the conventional methods introduced in Section~\ref{sec:baselines} and three learning-based alignment methods: \textit{Canonical Correlation Analysis} (\textbf{CCA})~\cite{faruqui2014cca, hotelling1936cca}, \textit{Embedding Converter} (\textbf{EC})~\cite{embeddingconverter}, and \textit{Drift-Adapter} (\textbf{DA})~\cite{drift-adapter}.\footnote{For learning-based methods, we follow the training-data construction and model architectures described in the original papers.}
In our experimental setting, the vectorDBs contain disjoint document subsets, so no candidate appears in more than one local result list.
Consequently, \textbf{RRF}~\cite{rrf} induces the same ranking as \textbf{AGG} \ra{up to deterministic tie-breaking}\footnote{\rb{We additionally evaluated \textbf{RRF} with 10\%, 50\%, and 90\% document overlap between two vectorDBs on Fever. RRF did not consistently improve over \textbf{HVMap}: its nDCG@10 ranged from 0.30 to 0.40, compared with 0.38--0.41 for \textbf{HVMap}. Although the gap narrows at high overlap, RRF provides no overall advantage.}}, while \(N(\mathbf{x})=1\) for every candidate in \textbf{CombMNZ}~\cite{combmnz}.
For \textbf{CombMNZ}, the local similarity scores are normalized using min--max normalization before fusion.

\textbf{CCA} learns linear projections of two embedding spaces into a shared latent space by maximizing the correlation between paired embeddings.
Each vectorDB is first queried using its corresponding query embedding to retrieve local candidates.
The queries and candidates are then projected into a 256-dimensional common space, globally reranked by common-space similarity, and truncated to the final top-\(K\) results.
\textbf{EC} converts stored vectors into a new embedding space using a trained mapping model. 
It can be viewed as a lower-cost variant of \textbf{MIG}, replacing corpus-wide re-embedding with vector conversion while retaining retrieval over a unified index. 
However, the converted vectors must still be re-indexed.
\textbf{DA} maps query vectors into each target vectorDB's embedding space, enabling local retrieval and aggregation without multi-model query embedding or corpus migration, at the cost of possible adapter-induced accuracy loss.

\subsubsection{Metrics and System Configuration}
\label{sec:metric_config}

To evaluate retrieval performance, we use two metrics: Recall@10 and normalized Discounted Cumulative Gain at rank 10 (nDCG@10)~\cite{a996db99c5094f0482d96dd6d94849c9}.
Recall@10 measures candidate coverage while largely abstracting away the ordering among retrieved documents, whereas nDCG@10 captures ranking quality.
Recall is also relevant to RAG pipelines in which all retrieved documents are supplied as context and their internal ordering is not explicitly used.
To determine whether observed performance differences are statistically significant, we apply the paired Wilcoxon signed-rank test over per-query results~\cite{wilcoxon1945individual}.
To evaluate computational efficiency, we estimate the total construction time, in hours, at a given data scale by extrapolating from measured runtime statistics.
For query throughput, we process 5,000 queries in batch, repeat each experiment five times, and report the average QPS. 
Error bars represent the standard deviation across the five runs.
\ra{Variance across independent mapping-function training runs was not assessed.}
All experiments, including construction-cost and query-throughput measurements, were conducted on a server equipped with an AMD EPYC 7763 CPU (128 cores, 3.5\,GHz), 1\,TB of RAM, and an NVIDIA A100 80\,GB GPU.
Unless otherwise noted, all experiments follow this default federated retrieval setup.

\begin{figure*}
  \centering
  \includegraphics[width=\linewidth]{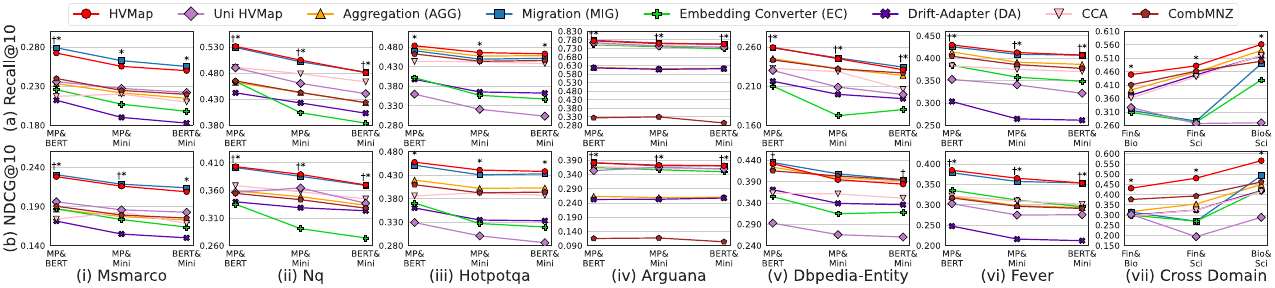}
  \caption{Recall@10 and nDCG@10 across six open-domain datasets (i--vi) and three cross-domain settings (vii). Each x-axis shows three embedding-model combinations. For \textbf{HVMap}, $\dagger$ denotes no significant difference from \textbf{MIG} ($p \geq 0.05$), and $*$ denotes a significant improvement over the strongest non-migration baseline ($p < 0.05$).}
  \label{fig:recall_comparison}
  \label{fig:ndcg_comparison}
\end{figure*}

\subsection{Base Case Evaluation: Two VectorDBs}
\label{sec:expr_basecase}
To demonstrate the effectiveness of \textbf{HVMap}, we first consider the base case of federated retrieval over two heterogeneous vectorDBs.

\begin{figure}[t!]
  \centering
  \includegraphics[width=1\linewidth]{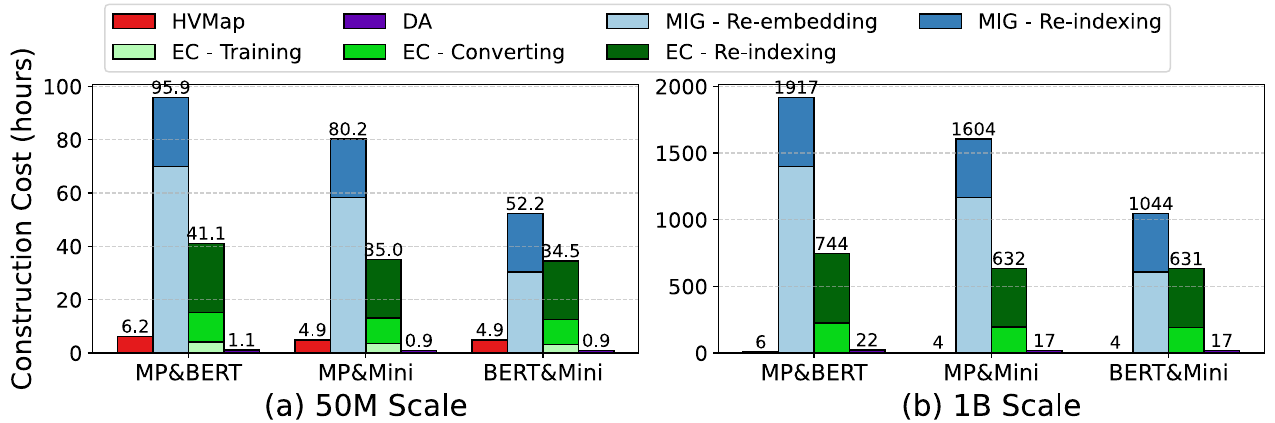}
  \caption{Decomposed construction costs across different embedding model combinations at the 50M and 1B scales.}
  \label{fig:construction_cost}
\end{figure}

\subsubsection{Retrieval Quality Evaluation}
\label{sec:basecase_recall}

Figure~\ref{fig:recall_comparison} compares \textbf{HVMap} with the baselines in terms of Recall@10 and nDCG@10 across six open-domain datasets and three cross-domain settings under three embedding-model combinations. 
\rb{We additionally evaluate Uni HVMap, a directional ablation that uses the same local candidate retrieval as \textbf{HVMap}, but ranks all candidates in one designated embedding space.} 
For \textbf{MIG} and \textbf{EC}, we report the average over the two possible migration directions to avoid favoring a particular target embedding space.

Among fusion-based methods, \textbf{AGG} and \textbf{CombMNZ} alternate in relative performance across settings. 
\ra{Although both exceed \textbf{HVMap} on Dbpedia-Entity for \textit{MP\&Mini} and \textit{BERT\&Mini} in terms of nDCG@10, and show competitive performance in the cross-domain settings, neither is robust across heterogeneous embedding-space configurations.}
\textbf{EC} largely follows \textbf{MIG} but incurs conversion error, while \textbf{DA} trades retrieval quality for lightweight query-side adapters. 
\textbf{CCA}'s linear shared-space projections are insufficient to capture the complex relationships between embedding spaces, resulting in lower retrieval accuracy than \textbf{HVMap}

Because \textbf{MIG} prioritizes retrieval accuracy through costly corpus-wide re-embedding and re-indexing, its comparison with \textbf{HVMap} is particularly important.
For Recall@10 on the open-domain datasets, \textbf{HVMap} outperforms \textbf{MIG} in 3 of the 18 settings, underperforms in 2, and shows no statistically significant difference in the remaining 13; nDCG@10 follows a similar pattern.
Moreover, \textbf{HVMap} significantly outperforms the strongest non-migration baseline in nearly all settings.

In cross-domain experiments, migration forces heterogeneous domains into one space, whereas fusion combines results without resolving cross-space incompatibility. 
\textbf{HVMap} preserves domain-specific vectorDBs and ranks candidates using bidirectional cross-space evidence, achieving the highest Recall@10 and nDCG@10 in all settings with significant gains over the strongest baseline. 
\rb{The lower performance of Uni HVMap further supports retaining native-space evidence through \textbf{HVMap}'s dual-space ranking.} 
The gains are especially large for \textit{Fin\&Bio} and \textit{Fin\&Sci}; because these datasets are excluded from mapping-function training, they also demonstrate generalization beyond the training distribution.
\rb{The impact of training-data composition is examined in Section~\ref{sec:mapping_ablation}.}

\subsubsection{Construction Cost Evaluation}

Figure~\ref{fig:construction_cost} presents a breakdown of the construction costs of methods with non-trivial setup overhead for 50M- and 1B-scale vectorDBs.
We exclude the fusion-based methods, \textbf{AGG} and \textbf{CombMNZ}, which require no additional system construction, as well as \textbf{CCA}, whose closed-form solution incurs negligible cost.
Because direct end-to-end measurement at the 1B scale was impractical, we extrapolate the embedding and indexing components from the 50M-scale measurements, as both scale linearly with the number of vectors under our fixed system configuration.
For \textbf{MIG}, the total cost consists of corpus re-embedding and re-indexing.
For \textbf{EC}, it includes converter training, vector conversion, and re-indexing.
For \textbf{HVMap} and \textbf{DA}, it corresponds to the training time of the mapping functions and adapters, respectively.

These results highlight a key motivation for federated retrieval by showing the extremely high construction cost of migration-based approaches, particularly \textbf{MIG}.
Both corpus-wide re-embedding and re-indexing grow proportionally with database size, while re-embedding cost also depends on the computational complexity of the embedding model.
\textbf{EC} substantially reduces this cost by replacing re-embedding with learned vector conversion, but still requires conversion of all stored vectors and subsequent re-indexing.
Accordingly, both components scale linearly with \(|D|\), yielding an overall construction cost of \(O(|D|)\).\footnote{The \textbf{MIG} results use a single GPU; multi-GPU embedding and GPU-accelerated indexing could reduce its absolute cost and narrow the gap with \textbf{HVMap}, but its construction cost would still scale as \(O(|D|)\).}

For \textbf{DA}, construction cost arises from adapter training.
Following the original setup, its training data are sampled from the target vectorDB and comprise approximately 2\% of the corpus.
Thus, its training cost also scales linearly with database size.
Although \textbf{DA} is about \(5\times\) cheaper than \textbf{HVMap} at the 50M scale, it becomes \(4\times\) more expensive at the 1B scale.

In contrast, \textbf{HVMap}'s construction cost is limited to training two bidirectional mapping functions for each pair of embedding spaces.
Because it uses a fixed-size training dataset rather than samples drawn from the target vectorDB, its cost remains effectively constant with respect to \(|D|\) under our training policy.
The primary factor affecting its training time is instead the embedding dimensionality, as reflected by the decreasing cost from \textit{MP\&BERT} to \textit{BERT\&Mini} in Figure~\ref{fig:construction_cost}.
Consequently, the cost gap between \textbf{HVMap} and the \(O(|D|)\) baselines widens as the database grows, making \textbf{HVMap} attractive for billion-scale deployments.

\subsubsection{Query Processing Time Evaluation}
\begin{figure}
  \centering
  \includegraphics[width=1\linewidth]{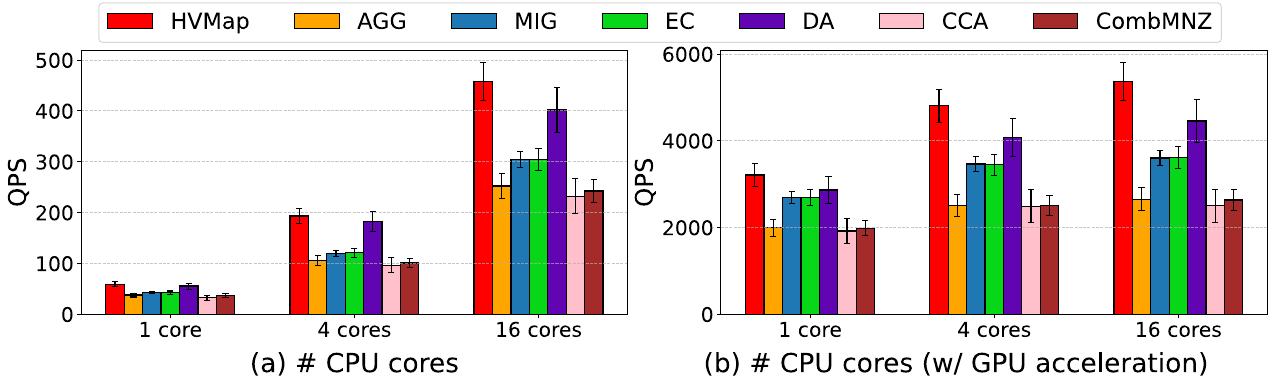}
    \caption{QPS comparison for federated retrieval over the MP\&Mini embedding-space pair on the NQ dataset, based on 5,000 queries. (a) reports results in a CPU-only environment, while (b) reports results with an A100 GPU.}
  \label{fig:QPS_CS}
\end{figure}

We evaluate query throughput (QPS) on Nq using the \textit{MP\&Mini} embedding pair under CPU-only and GPU-enabled settings with 1, 4, and 16 CPU cores.
\textbf{MIG} and \textbf{EC} perform embedding and search; \textbf{AGG}, \textbf{CombMNZ}, and \textbf{CCA} use two query embeddings followed by local retrieval and fusion or reranking; \textbf{HVMap} performs embedding, search, mapping, and ranking; and \textbf{DA} uses embedding, adapter inference, search, and aggregation.
For \textbf{MIG} and \textbf{EC}, we migrate to the stronger \textit{MP} space, while \textbf{DA} results are averaged over both adapter directions.
Figures~\ref{fig:QPS_CS}(a) and (b) report the corresponding results.

Methods that use two embedding models exhibit relatively low throughput.
\textbf{AGG} and \textbf{CombMNZ} embed each query with both participating models before local retrieval and fusion.
\textbf{CCA} incurs the same dual-embedding overhead and additionally projects the queries and candidates into a common space, resulting in slightly lower QPS.
Because query embedding is the primary bottleneck, differences in fusion or reranking cost have a comparatively minor impact on overall throughput.

In contrast, \textbf{MIG}, \textbf{EC}, \textbf{DA}, and \textbf{HVMap} require only one query embedding.
\textbf{MIG} and \textbf{EC} search a unified vectorDB, but their use of the heavier \textit{MP} embedding model limits throughput.
\textbf{DA} replaces additional query embeddings with lightweight adapters and therefore achieves substantially higher QPS than the dual-embedding methods.
\textbf{HVMap} further benefits from using the lightweight \textit{Mini} model and achieves the highest QPS across all CPU-only and GPU-enabled configurations.
This advantage is primarily driven by differences in embedding cost.
While \textbf{MIG} and \textbf{EC} embed queries in the \textit{MP} space, \textbf{HVMap} can use the lightweight \textit{Mini} model, which is 5--7$\times$ faster, while preserving retrieval quality through bidirectional mapping.
The mapping component is highly GPU-accelerated and supports large batches, incurring only about 0.2\% of the embedding cost.
After embedding, vector search becomes the dominant component; increasing the number of CPU cores therefore improves throughput further, particularly for \textbf{HVMap}, which searches multiple vectorDBs in parallel.
A detailed component-wise latency breakdown is provided in Section~\ref{sec:scale_system}.

\begin{figure*}[t!]
  \centering
  \includegraphics[width=0.9\linewidth]{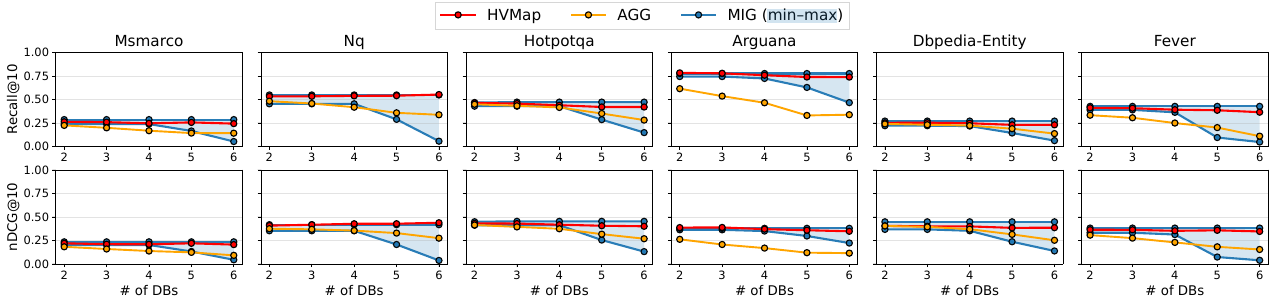}
  \caption{
    Recall@10 (top) and nDCG@10 (bottom) as the number of vectorDBs increases from two to six across six open-domain datasets. 
    VectorDBs are added in the order of \textit{Mini}, \textit{MP}, \textit{BERT}, \textit{RoBERTa}, \textit{Sentence-T5}, and \textit{LaBSE}. 
    For \textbf{HVMap}, \textit{Mini} is used as the hub space. 
    For \textbf{MIG}, the shaded region denotes the min--max performance across all migration directions.
}
\label{fig:comparison_scale}
\end{figure*}

\subsection{Evaluation at Scale: Multiple VectorDBs}
\label{sec:scale_expr}
In this section, we extend the evaluation to settings with up to six vectorDBs to analyze the scale-out behavior of each method.
We report \textbf{MIG} as the representative migration-based method and \textbf{AGG} as the representative fusion-based method, since extending the remaining baselines to multi-vectorDB federation would require additional design choices beyond their original scope and could confound the comparison.

\begin{figure}
  \centering
  \includegraphics[width=1\linewidth]{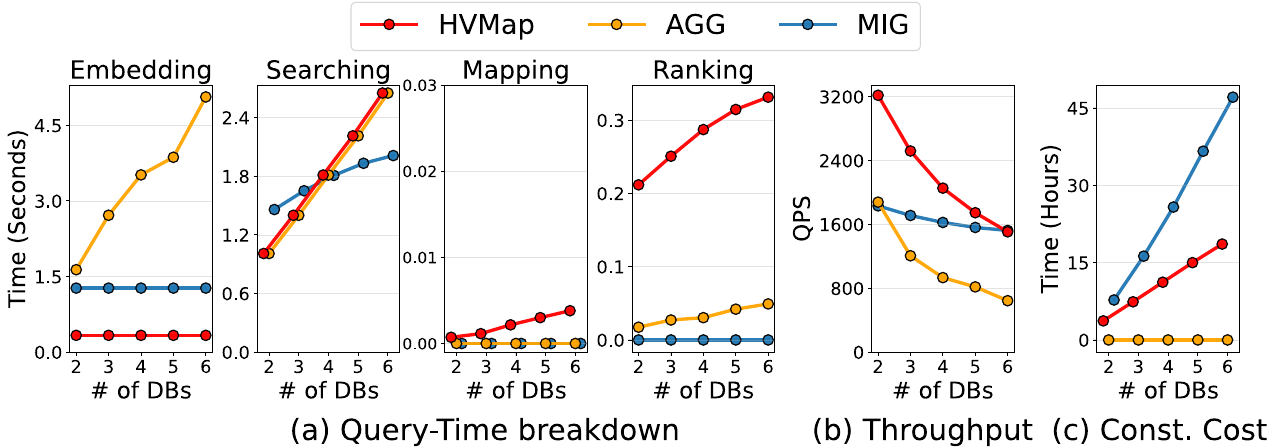}
    \caption{Comparison of (a) query-time breakdown, (b) throughput, and (c) construction cost across different methods as the number of vectorDBs scales from 2 to 6.}
  \label{fig:scale_qps_cost}
\end{figure}

\subsubsection{Retrieval Robustness}

Figure~\ref{fig:comparison_scale} shows Recall@10 and nDCG@10 as the number of vectorDBs increases from two to six across six open-domain datasets.
The vectorDBs are added in the order \textit{Mini}, \textit{MP}, \textit{BERT}, \textit{RoBERTa}, \textit{Sentence-T5}, and \textit{LaBSE}\footnote{In our experiments, retrieval effectiveness tended to decrease in the order of \textit{MP}, \textit{BERT}, \textit{RoBERTa}, \textit{Sentence-T5}, and \textit{LaBSE}. As weaker models are added later, the minimum accuracy of \textbf{MIG} tends to decrease with the number of vectorDBs.}.
For \textbf{MIG}, performance is reported as the minimum and maximum over migration directions, since the best unified target space is generally unknown in advance.
For \textbf{HVMap}, the lightweight \textit{Mini} space is selected as the hub to maximize query throughput, trading a small amount of accuracy for substantially higher efficiency.
The results show that the lower bound of \textbf{MIG} decreases markedly as more vectorDBs are added, because migrating all vectors into weaker embedding spaces can degrade retrieval quality.
\textbf{AGG} also declines with scale, as its assumption that relevant documents are evenly distributed across vectorDBs becomes increasingly inaccurate.
In contrast, \textbf{HVMap} maintains stable performance as the federation grows.
Its hub-based structure limits mapping-error propagation to at most two steps, while scoring candidates in all embedding spaces allows it to exploit the strengths of every model.
Consequently, \textbf{HVMap} remains close to the best migration directions of \textbf{MIG} across datasets, demonstrating robust retrieval quality under scale-out.

\begin{table}[t]
\centering
\caption{Asymptotic cost comparison of construction and query-time breakdown, where \(N\) denotes the number of vectorDBs and \(|D|\) denotes the size of each database.}
\footnotesize
\setlength{\tabcolsep}{6pt}
\renewcommand{\arraystretch}{1.2}
\begin{adjustbox}{width=\linewidth}
\begin{tabular}{lccccc}
\toprule
& \textbf{Construction} & \multicolumn{4}{c}{\textbf{Query-Time Breakdown}} \\
\cmidrule(lr){2-2} \cmidrule(lr){3-6}
\textbf{Method} & \textbf{Cost} & \textbf{Embedding} & \textbf{Searching} & \textbf{Mapping} & \textbf{Ranking} \\
\midrule
\textbf{HVMap}
& $O(N)$
& $O(1)$ 
& $O(N \log |D|)$ 
& $O(N)$ 
& $O(N^2)$  \\ 
\textbf{AGG}
& -- 
& $O(N)$ 
& $O(N \log |D|)$ 
& --
& $O(N)$ \\
\textbf{MIG}
& $O(N|D|)$
& $O(1)$ 
& $O(\log (N|D|))$ 
& --
& -- \\
\bottomrule
\end{tabular}
\end{adjustbox}
\label{tab:asymtotic}
\end{table}

\subsubsection{System Scalability}
\label{sec:scale_system}
We next evaluate system scalability by examining whether execution and construction costs remain practical.
Heterogeneous federation avoids migration and reconstruction but requires queries to span multiple vectorDBs and embedding spaces.
Figure~\ref{fig:scale_qps_cost} and Table~\ref{tab:asymtotic} report query-time breakdown, modeled throughput, and construction cost as the number of vectorDBs increases from 2 to 6 while the total corpus size remains fixed.
Accordingly, each vectorDB becomes smaller as \(N\) increases, isolating the effect of federation scale from corpus size.
\textbf{HVMap} scales favorably because its hub-based architecture localizes compatibility costs.
Query embedding is constant in \(N\), since only the hub model encodes the query.
Mapping grows linearly with \(N\) but remains small through batched transformations, while ranking is \(O(N^2)\) because each embedding space scores candidates from all vectorDBs.
This overhead remains modest in practice because \(K\) is fixed, \(N\) is small, and scoring uses batched GPU operations.
Thus, search dominates under scale-out: non-migration-based federation must search \(N\) separate vectorDBs, yielding the \(O(N\log |D|)\) cost in Table~\ref{tab:asymtotic}.
\textbf{HVMap} does not eliminate this cost, but keeps compatibility overhead lightweight.
Its construction cost is linear in \(N\), requiring \(2(N-1)\) mappings, and depends on the number of spaces.

\textbf{AGG} scales much less gracefully. 
Because it lacks an explicit mechanism for cross-space compatibility, it must pay repeatedly at query time: one query embedding per vectorDB, one search per vectorDB, and then a ranking step over the combined candidates. 
Its embedding and search costs therefore both grow linearly with \(N\), while its retrieval quality simultaneously deteriorates. 
This is an especially unfavorable scaling pattern because the method becomes both slower and less accurate as federation expands.
\textbf{MIG} follows the opposite philosophy. 
By paying the one-time cost of full migration and re-indexing, it eliminates cross-space operations during query processing altogether. 
Once all vectors have been integrated into a single MP-space vectorDB, throughput is governed only by HNSW search over the unified corpus, producing the \(O(\log (N|D|))\) search-time behavior in Table~\ref{tab:asymtotic}. 
This gives \textbf{MIG} the most favorable online scaling among the compared methods. 
However, this advantage is obtained at construction cost \(O(N|D|)\), since full migration and re-indexing must be performed over the entire corpus.

Figure~\ref{fig:scale_qps_cost} shows that \textbf{HVMap} and \textbf{MIG} shift the system bottleneck differently rather than one dominating every axis.
\textbf{MIG} improves online search scaling by consolidating all vectors into one database, whereas \textbf{HVMap} preserves vectorDB autonomy with fundamentally lower construction cost and small compatibility overhead relative to multi-database search.
At \(N=6\), \textbf{HVMap}'s throughput approaches that of \textbf{MIG}, while a substantial construction cost gap remains despite only \(2.6\)M vectors per vectorDB and \(16\)M in total.
This gap should widen as \(N\) and \(|D|\) grow because \textbf{MIG} scales with total corpus size, whereas \textbf{HVMap} does not.

\begin{table}[t]
\centering
\caption{nDCG@10 of \textbf{HVMap} under different mapping objectives.
Bold values indicate the best result in each setting, and $*$ denotes a significant improvement of Fuzzy CL over the strongest alternative under a Wilcoxon signed-rank test.}
\footnotesize
\label{tab:loss_joint}
\setlength{\tabcolsep}{3pt}
\renewcommand{\arraystretch}{1.15}
\begin{adjustbox}{width=\linewidth}
\begin{tabular}{lcccccccccccc}
\toprule
\multirow{2}{*}{\textbf{Dataset}}
& \multicolumn{4}{c}{\textbf{MP\&BERT}}
& \multicolumn{4}{c}{\textbf{MP\&Mini}}
& \multicolumn{4}{c}{\textbf{BERT\&Mini}} \\
\cmidrule(lr){2-5}
\cmidrule(lr){6-9}
\cmidrule(lr){10-13}
& Fuzzy CL & CL & MSE & RKD+MSE
& Fuzzy CL & CL & MSE & RKD+MSE
& Fuzzy CL & CL & MSE & RKD+MSE \\
\midrule

Msmarco
& \textbf{0.2296}  & \ra{0.2219} & \ra{0.2120} &  {0.2272}
& \textbf{0.2157}* & \ra{0.2062} & \ra{0.1892} &  {0.2115}
& \textbf{0.2086}* & \ra{0.2027} & \ra{0.1797} &  {0.2037} \\

Nq
& \textbf{0.4027}  & \ra{0.3528} & \ra{0.3318} &  {0.3998}
& \textbf{0.3886}* & \ra{0.3584} & \ra{0.3121} &  {0.3830}
& \textbf{0.3694}* & \ra{0.3592} & \ra{0.3029} &  {0.3642} \\

Hotpotqa
& \textbf{0.4572}* & \ra{0.3442} & \ra{0.3134} &  {0.4345}
& \textbf{0.4405}* & \ra{0.3319} & \ra{0.3415} &  {0.4187}
& \textbf{0.4376}* & \ra{0.3290} & \ra{0.2690} &  {0.4159} \\

Arguana
& \textbf{0.3798}  & \ra{0.3719} & \ra{0.3673} &  {0.3794}
& \textbf{0.3721} & \ra{0.3591} & \ra{0.3576} &  {0.3690}
& \textbf{0.3700}* & \ra{0.3517} & \ra{0.3499} &  {0.3672} \\

\makecell[l]{Dbpedia\\Entity}
& \textbf{0.4315}* & \ra{0.3126} & \ra{0.2918} &  {0.4079}
& \textbf{0.3940}  & \ra{0.2978} & \ra{0.2662} &  {0.3939}
& \textbf{0.3835}* & \ra{0.3002} & \ra{0.2788} &  {0.3731} \\

Fever
& \textbf{0.3841}* & \ra{0.3477} & \ra{0.3311} &  {0.3743}
& \textbf{0.3645}* & \ra{0.3223} & \ra{0.3029} &  {0.3586}
& \textbf{0.3529}* & \ra{0.3097} & \ra{0.3071} &  {0.3427} \\

\bottomrule
\end{tabular}
\end{adjustbox}
\end{table}

\begin{figure}[t]
    \centering
    \includegraphics[width=\linewidth]{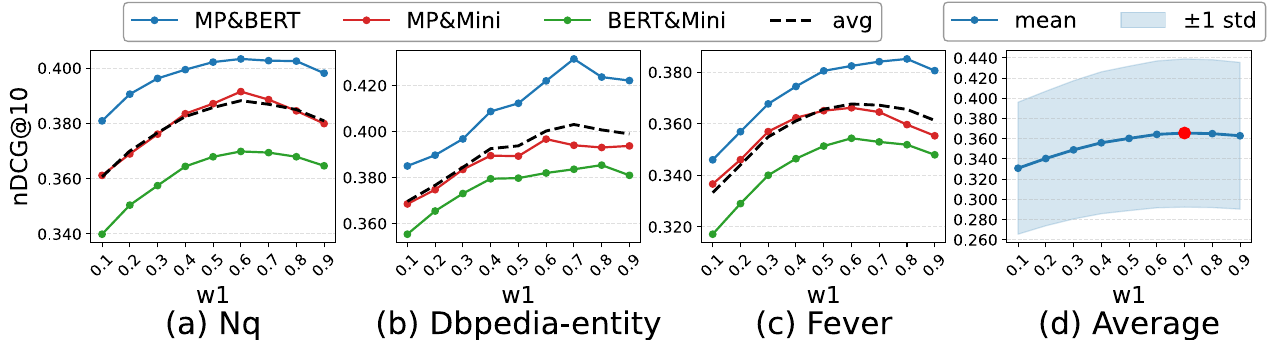}
    \caption{
    Effect of the $w_1$ on retrieval accuracy of HVMap.
    The dashed line denotes the average across model pairs, and (d) reports the average over all six datasets and model pairs.
    }
    \label{fig:score_weight}
\end{figure}

\subsection{Ablation Studies}

\subsubsection{Effectiveness of Fuzzy CL}
\label{sec:loss_compare}

To evaluate the fuzzy contrastive loss, we compare \textbf{HVMap} trained with Fuzzy CL, CL, MSE, and RKD combined with MSE \ra{(hereafter, RKD+MSE)}, following the original framework.
Table~\ref{tab:loss_joint} reports nDCG@10 across six open-domain datasets and three embedding-space federation scenarios.
Fuzzy CL achieves the highest nDCG@10 in all 18 settings and significantly outperforms the strongest alternative in 13, while RKD+MSE consistently improves over MSE or CL alone.
\ra{More importantly, the significant advantage over RKD+MSE in 13 of 18 settings indicates that Fuzzy CL is not merely a combination of pointwise and relational losses.}
Instead, it provides a unified, retrieval-aware objective that learns the cross-space score structure required by an unchanged target vectorDB, yielding more precise alignment than conventional CL or RKD+MSE.

\subsubsection{Analysis of SCORE$_2$ Weighting}
\label{sec:weight_study}

We vary the native-space weight $w_1$ from 0.1 to 0.9 to optimize the weighted SCORE$_2$ function for HVMap.
Figure~\ref{fig:score_weight} reports the results on three representative datasets and the overall average across all six datasets and embedding-model pairs. 
The best average performance is obtained at $w_1=0.7$, with stable performance between 0.5 and 0.8. 
This indicates that native-space similarity should receive slightly greater weight because it is unaffected by mapping error, while mapped-space similarity provides complementary cross-space calibration.

\begin{figure}[t]
    \centering
    \includegraphics[width=\linewidth]{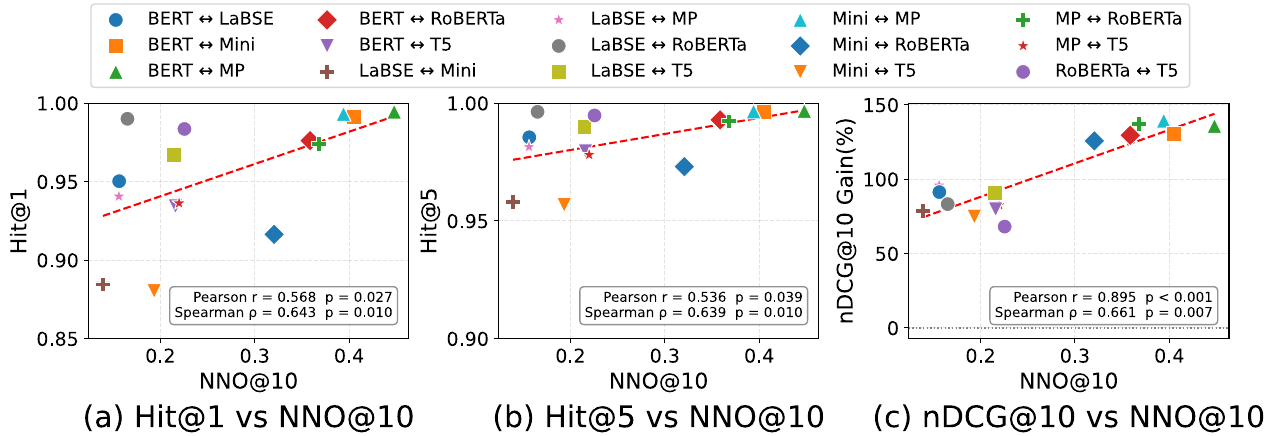}
    \caption{
    Relationship between embedding-space similarity, mapping quality, and retrieval gain.
    (a)--(b) compare NNO@10 with Hit@1 and Hit@5, while (c) reports the nDCG@10 gain over an untrained dummy mapping.
    }
    \label{fig:nno}
\end{figure}

\subsubsection{Embedding-Space Similarity and Mapping Quality}
\label{sec:nno}
We analyze how embedding-space similarity affects mapping quality and retrieval accuracy.
We measure similarity using Nearest Neighbor Overlap (NNO)~\cite{nno,huh2024platonic}, which quantifies overlap between local neighborhoods induced by two embedding models.
Mapping quality is measured by Hit@$K$, indicating whether a mapped embedding retrieves its paired target within the top-$K$ neighbors.
Figures~\ref{fig:nno}(a) and (b) show that higher NNO generally corresponds to higher Hit@1 and Hit@5, indicating that structurally similar spaces are easier to align.
However, even low-NNO pairs exceed 95\% Hit@5, showing that low similarity does not necessarily cause mapping failure.
Figure~\ref{fig:nno}(c) further shows that higher NNO generally leads to larger nDCG@10 gains over an untrained dummy mapping.
However, \textbf{HVMap} improves retrieval quality by at least 68\% for all evaluated pairs, indicating that NNO is a useful diagnostic of mapping difficulty while fuzzy contrastive mapping remains effective even across structurally dissimilar spaces.




\subsubsection{Impact of Mapping Capacity and Training Data}
\label{sec:mapping_ablation}

\begin{figure}[t]
    \centering
    \includegraphics[width=\linewidth]{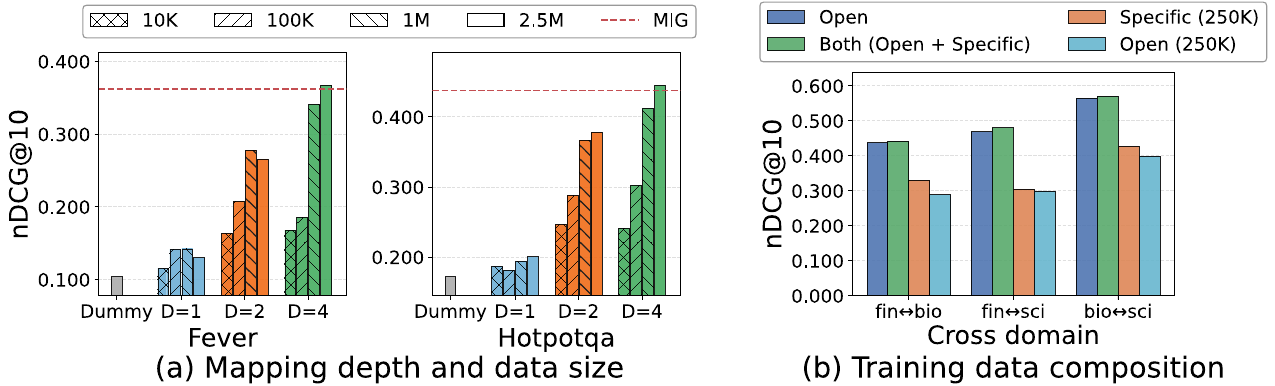}
    \caption{Impact of mapping capacity and training data.
    (a) Mapping depth and training-data size; red dashed lines denotes \textbf{MIG}.
    \rb{(b) Cross-domain results for training-data compositions; 250K denotes a size-matched corpus.}}
    
    \label{fig:mapping_ablation}
\end{figure}

Figure~\ref{fig:mapping_ablation}(a) shows the effect of mapping depth and paired training-data size on \textbf{HVMap}.
One-layer mapping performs similarly to the dummy baseline, whereas greater depth and more training data generally improve retrieval quality.
A 4-layer MLP trained with 2.5M paired samples is the smallest tested configuration that reaches or exceeds \textbf{MIG} on both datasets shown.
\rb{Figure~\ref{fig:mapping_ablation}(b) further evaluates training-corpus composition for cross-domain federation.
With 250K samples, domain-specific data slightly outperforms open-domain data.
However, a larger open-domain corpus substantially improves performance, and adding domain-specific data provides only limited benefit when sufficient open-domain training data are available.
Overall, effective cross-space mapping requires sufficient nonlinear capacity and paired training data, with training-data scale and diversity having a greater impact as data availability grows.}


\subsubsection{Retrieval Sensitivity Analysis}
\label{sec:tuned}

\begin{figure}[t]
  \centering
  \includegraphics[width=\linewidth]{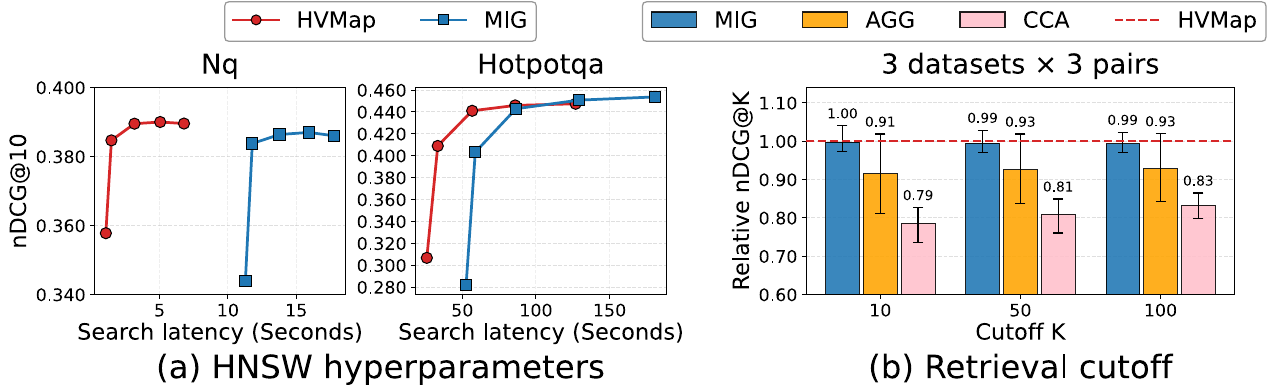}
  \caption{Retrieval sensitivity analysis.
    (a) nDCG@10 versus latency under HNSW configurations.
    \rb{(b) nDCG@$K$ relative to HVMap; mean over 9 settings, bars show min--max.}}
  \label{fig:tuning}
\end{figure}

Section~\ref{sec:expr_basecase} uses a common default HNSW configuration.
\ra{On the open-domain Nq and Hotpotqa datasets,} we vary $(M,\texttt{efC})$ over $(8,64)$, $(16,128)$, $(32,256)$, $(48,384)$, and $(64,512)$ for \textbf{MIG} and \textbf{HVMap}.
Figure~\ref{fig:tuning}(a) reports nDCG@10 against cumulative vectorDB search time; both methods benefit from tuning, while \textbf{HVMap} retains a favorable latency--accuracy trade-off, indicating robustness to HNSW hyperparameter choices.


\rb{We also vary the retrieval cutoff $K$ over 10, 50, and 100 across Dbpedia-Entity, Fever, and Hotpotqa.
The relative performance trends remain broadly consistent across all cutoffs, showing that the choice of $K$ does not materially alter the comparative results.}

\subsubsection{\rb{Robustness to Skewed Relevance}}
\label{sec:skewed_relevance}
\rb{We compare a skewed Nq split, in which all relevant documents reside in one vectorDB, with a balanced split.
For \textit{MP\&BERT}, \textit{MP\&Mini}, and \textit{BERT\&Mini}, \textbf{HVMap} achieves nDCG@10 of 0.403 versus 0.403, 0.389 versus 0.386, and 0.369 versus 0.370 in the skewed and balanced settings, respectively.
These results show that retrieving up to $K$ candidates per source does not impose an equal final-result allocation: global reranking allows one vectorDB to contribute all $K$ results when appropriate.}

\section{Limitation \& Discussion}
\label{sec:lim_dis}

\noindent\textbf{Deployment assumptions.}
HVMap avoids corpus-wide migration, but it requires offline training of mapping functions.
This training uses paired embeddings generated by applying the participating embedding models to the same auxiliary samples.
Accordingly, HVMap assumes access to the embedding models or their APIs during construction.
Deployment may be difficult when such access is unavailable due to privacy, licensing, or API restrictions. 

\noindent
\noindent\textbf{Score weighting and calibration.}
HVMap combines native-space and mapped-space similarities using a weighted SCORE function.
\ra{Our analysis supports $w_{1}=0.7$ for the two-vectorDB setting, but a fixed weight may not be optimal for all deployments.
In larger federations, suitable weights may depend on the number of vectorDBs, the hub model, mapping quality, and embedding-space similarity.}

\noindent
\noindent\textbf{Cross-modal federation.}
\textbf{HVMap} is not inherently limited to text embeddings and could potentially support federation across modalities while preserving modality-specific representation spaces.
However, cross-modal mapping requires paired data, such as image--caption pairs, and may benefit from modality-specific mapping architectures or training objectives.
We therefore leave general multimodal federation as future work.


\section{Related Work}
\textbf{Embedding Compatibility and vectorDB Integration.}
Prior work has extensively studied vector compatibility under embedding model updates and heterogeneous representation spaces~\cite{jaeckle2023fastfill,Shen_2020_CVPR,hu2022learning,ramanujan2022forward,park2019relational}.
Most methods enforce compatibility during training or project embeddings into a shared space, primarily treating the problem from a representation-learning perspective.
They therefore do not explicitly address system-level concerns in federating independently constructed vectorDBs, including construction cost, query throughput, and scalability.
More recently, integration-oriented methods have used locally isometric mappings to combine heterogeneous vectorDBs~\cite{integration}.
However, they target database integration rather than query-time federation and mainly report post-integration retrieval effectiveness, leaving integration overhead and scalability largely unexplored.

\noindent\textbf{Federated Retrieval Systems.}
Federated databases have long been studied as a classical topic in database systems, with the goal of integrating distinct database systems while preserving local autonomy~\cite{10.1145/96602.96604, 107449}.
Prior work on federated relational databases has focused on issues such as query optimization~\cite{10.5555/645923.670995}, schema integration~\cite{10.1145/27633.27634, 10.5555/645926.671677}, and consistency management~\cite{10.1007/BF01228709}.
More recently, several studies have explored federated architectures for vector retrieval~\cite{zhu2024fedsq,zhao2024frag}.
These works primarily focus on distributed execution, data privacy, or infrastructure-level concerns, rather than the compatibility problem caused by heterogeneous embedding spaces.
In contrast, our work treats retrieval compatibility as a first-class systems problem and specifically targets federated retrieval across vectorDBs built on incompatible embedding models.

\section{CONCLUSION}

We presented HVMap, a lightweight compatibility layer for federated retrieval across heterogeneous vectorDBs with incompatible embedding spaces. 
By learning bidirectional mappings with fuzzy contrastive loss, HVMap avoids costly corpus re-embedding and index reconstruction while maintaining strong retrieval performance.
Our results show that heterogeneous vectorDB federation is a systems trade-off among retrieval quality, construction cost, and query throughput, and that HVMap provides a practical solution in this design space. 
Its hub-based architecture enables scalable retrieval across multiple vectorDBs without quadratic mapping complexity.

\begin{acks}
 This work was supported in part by the National Research Foundation of Korea (NRF) grants (RS-2025-00560762, RS-2024-00414981), the Institute of Information \& Communications Technology Planning \& Evaluation (IITP) grants (RS-2024-00454666, RS-2025-25442338, RS-2026-25518317, RS-2026-25551943, RS-2021-II211343), Samsung Electronics Co., Ltd. (IO251215-14535-01), and the Ministry of Science and ICT under the Advanced GPU Utilization Support Program. The authors thank Samsung MAX Lab for providing research infrastructure. J. Do is with ASRI, Seoul National University.
\end{acks}



\balance
\bibliographystyle{ACM-Reference-Format}
\bibliography{sample}

\end{document}